%% file: main.tex
\documentclass{article}
\usepackage[utf8]{inputenc}
\usepackage{natbib}
\usepackage[inline,shortlabels]{enumitem}
\usepackage{booktabs}
\usepackage{multirow}
\usepackage[a4paper,margin=1in]{geometry}
\usepackage{url}
\usepackage{tablefootnote}
\usepackage{graphicx}
\usepackage{float}
\usepackage[justification=centering]{caption}
\usepackage{xcolor}
\usepackage{soul}
\usepackage{array}
\usepackage{threeparttable}
\usepackage{tabularx}
\usepackage{makecell}
\usepackage{subcaption}
\usepackage{caption}
\usepackage{authblk}
\usepackage{bold-extra}

        \usepackage{etoolbox} %
        \AtBeginEnvironment{tabularx}{\setlist[itemize, 1]{wide, leftmargin=*, itemsep=0pt, before=\vspace{-\dimexpr\baselineskip +1 \partopsep}, after=\vspace{-\baselineskip}}}

        \usepackage{etoolbox} %
        \AtBeginEnvironment{tabularx}{\setlist[enumerate, 1]{itemsep=0pt, before=\vspace{-\dimexpr\baselineskip +1 \partopsep}, after=\vspace{-\baselineskip}}}
        
\newcommand{\quotes}[1]{``#1''}

\title{The Impact of Annotation Guidelines and Annotated Data on Extracting App Features from App Reviews}

\author{Faiz Ali Shah}
\author{Kairit Sirts}
\author{Dietmar Pfahl}
\affil{Institute of Computer Science, University of Tartu, \\ J. Liivi 2, 50409, Tartu, Estonia.}

\begin{document}
\date{}
\maketitle

\section*{Abstract}

The quality of automatic app feature extraction from app reviews depends on various aspects, e.g. the feature extraction method, training and evaluation datasets, evaluation method etc. Annotation guidelines used to guide the annotation of training and evaluation datasets can have a considerable impact to the quality of the whole system but it is one of the aspects that has been commonly overlooked. In this study, we explore the effects of annotation guidelines to the quality of app feature extraction. As a main result, we propose several changes to the existing annotation guidelines with a goal of making the extracted app features more useful and informative to the app developers. We test the proposed changes via simulating the application of the new annotation guidelines and then evaluating the performance of the supervised machine learning models trained on datasets annotated with initial and simulated annotation guidelines. While the overall performance of automatic app feature extraction remains the same as compared to the model trained on the dataset with initial annotations, the features extracted by the model trained on the dataset with simulated new annotations are less noisy and more informative to the app developers.  Secondly, we are interested in what kind of annotated training data is necessary for training an automatic app feature extraction model. In particular, we explore whether the training set should contain annotated app reviews from those apps/app categories on which the model is subsequently planned to be applied, or is it sufficient to have annotated app reviews from any app available for training, even when these apps are from very different categories compared to the test app. Our experiments show that having annotated training reviews from the test app is not necessary although including them into training set helps to improve recall. Furthermore, we test whether augmenting the training set with annotated product reviews helps to improve the performance of app feature extraction. We find that the models trained on augmented training set lead to improved recall but at the cost of the drop in precision. \\
\newline
\noindent
\textbf{Keywords:}  app features extraction; opinion mining; sentiment analysis; sequence modeling

\input{introduction}
\input{related_work}
\input{design_variables}

\input{study_design}
\input{results}

\input{discussion}
\input{conclusion}

\bibliographystyle{apalike}
\bibliography{references}

\appendix
\input{appendix_A}

\input{appendix_B}
\input{appendix_C}

\end{document}

%% file: introduction.tex
\section{Introduction}
\label{sec:introduction}

App marketplaces provide app users a channel to submit feedback in the form of a review. Users in these reviews provide valuable information such as feature requests, bug reports, user experience, and evaluation of app features \citep{pagano2013}. The analysis of opinions expressed about different features of an app in user reviews offers opportunities and insights to both app users and app developers.  For app developers, it is useful to monitor the ``health'' of app features in the context of release planning \citep{maalej2016} as well as to evaluate product competitiveness and quality \citep{shah2016}. From the users' perspective, such information helps in deciding which app to select from a wide range of competing apps. Both Apple's App Store and Google's Play Store receive enormous amounts of reviews every day making a manual analysis infeasible and demanding automated methods. One standard approach towards this goal is to generate sentiment summaries of a product at feature-level involving two steps \citep{zhang2014}: 
\begin{enumerate*}[label=\arabic*)]
\item identification of app features (also called \emph{aspect terms} or \emph{opinion targets} in the opinion mining literature) in user reviews, and 
\item determination and aggregation of sentiments expressed on product features identified in the previous step.
\end{enumerate*}

Since identifying product features in user reviews is a crucial step for generating sentiment summaries, several prior studies on app review analysis have exclusively focused on this step. Previous work on aspect extraction from app reviews includes using unsupervised topic modeling \citep{guzman2014} and an approach called \textsc{SAFE} (Simple Approach for Feature Extraction) based on manually extracted rules \citep{johann2017}. The performance reported with these two approaches is summarized in the first two rows of Table~\ref{tab:intro}. Although these results seem to indicate that the topic modeling and the SAFE approach are complementary with regards to precision and recall, they are not directly comparable because the authors used different sets of app reviews, different annotation guidelines and different performance evaluation methods. \citet{johann2017} compared the performance of a version of the topic modeling approach proposed by \citet{guzman2014} on their set of app reviews using the same annotation guidelines and evaluation procedure as in the \textsc{SAFE} approach. These results, shown in the 3rd row of Table~\ref{tab:intro}, indicate that the performance of the topic modeling approach is much lower than reported in the original paper and also worse than that of the \textsc{SAFE} approach, in particular with regards to recall. Another recent study for extracting features from app reviews, conducted on German language reviews \citep{sanger2016}, used a supervised machine learning method of Conditional Random Fields (CRF) \citep{lafferty2001}. Previously, CRF method has been successfully applied to extract features from \textsc{Laptop} and \textsc{Restaurant} reviews \citep{toh2014,liu2015}, which serve as standard aspect extraction benchmark datasets in sentiment analysis community \citep{pontiki2014}. The results of \citet{sanger2016}, shown in the last row of Table~\ref{tab:intro},
suggest that supervised machine learning methods might have an advantage over unsupervised and rule-based methods (such as SAFE) also for extracting features from English app reviews, given that annotated training data is available. 


\begin{table}
\centering
\begin{tabular}{lccc}
\toprule
\bf Approach & \bf Precision & \bf Recall & \bf F-score \\
\midrule
\bf Topic modeling (1) & \multirow{2}{*}{0.58} & \multirow{2}{*}{0.52} & \multirow{2}{*}{0.55} \\
\citep{guzman2014} & & \\
\textbf{SAFE} \citep{johann2017} & 0.24 & 0.71 & 0.36 \\
\bf Topic modeling (2) & \multirow{2}{*}{0.22} & \multirow{2}{*}{0.28} & \multirow{2}{*}{0.24} \\
\citep{johann2017} & & & \\
\textbf{CRF} \citep{sanger2016} &  0.69 & 0.56 & 0.62 \\
\bottomrule
\end{tabular}
\caption{Performance obtained with different approaches to extract features from app reviews}
\label{tab:intro}
\end{table}

There are several questions these previous studies fail to answer when developing systems for automatically extracting features from app reviews. The first question is related to the annotation of app features: Which word or sequence of words in a review constitutes a feature of an app? Training a feature extraction system using supervised machine learning methods requires a training set where all feature instances are annotated\footnote{Note that we use the words \emph{annotated} and \emph{labeled} as synonyms in this article.}. Unsupervised or rule-based systems do not need an annotated training set but they need an annotated test set to evaluate how well the system performs. Clearly, the exact annotation procedure, operationalized via annotation guidelines, has potentially a large effect on both the evaluation results and the usefulness of those results to app developers. This motivates our first research question: \\

\noindent
\textbf{RQ1:} To what extent is automatic feature extraction from app reviews sensitive to the used annotation guidelines? \\

\noindent
To study this research question, we use two available app review datasets that are both annotated with app features using different annotation guidelines: 
\begin{enumerate*}[label=\arabic*)]
\item English app review data contributed by \citet{guzman2014}, which includes annotated reviews of seven apps from App Store and Play Store (\textsc{Guzman} dataset) and
\item German app review data published by \citet{sanger2016}, which contains annotated reviews from eleven app categories from Play Store (\textsc{Sanger} dataset).\footnote{The particular apps from which the reviews are taken are not known.} 
\end{enumerate*}
Furthermore, we employed two undergraduate students as annotators. Each of them annotated independently 500 reviews of the seven apps contained in the \textsc{Guzman} dataset\footnote{Although the reviews our students annotated are not the same as the ones used in \textsc{Guzman} dataset.} following the annotation guidelines proposed by \citet{sanger2016}. 
Because the inter-annotator agreement between the two annotators on the newly annotated dataset is low (Dice index = 0.28), we treat the annotations of the two annotators as two different datasets.
We train and evaluate supervised CRF models on all datasets, i.e., the \textsc{Guzman} dataset, the \textsc{Sanger} dataset, and the datasets annotated by our student annotators.
We use Cross-Category Validation procedure for training and evaluation, i.e., 
we assume that our dataset consists of app reviews from several different categories (such as Games, Productivity or Sports).
Each app category contains reviews from one or more apps belonging into that category. Then we select to hold out one app category and train the model on the app reviews of all other categories. Finally, we test on the app reviews of the held-out app category. This procedure is repeated until all categories have been held out in turn. Then we use the average feature extraction performance of these models as a proxy for evaluating the goodness of the annotation guidelines. We simulate several changes in annotation guidelines and assessed their effect using the performance of the predictive CRF modeling. Using this procedure, we are able to propose several changes to the app feature annotation guidelines that improve the quality of the annotated app reviews for both training and evaluation purposes.

Another important practical question is how much annotated data one needs to train a generalizable and reliable model. The functional features 
can vary across different apps
while many apps may share a common set of non-functional features. Does this common set of app features provide enough information for the model so that it will generalize well enough also to the reviews of the apps for which no annotated training data is available? Or is it necessary to restrict oneself to app reviews of the same app category, and is it even necessary to annotate a certain amount of reviews for every new app that we want to analyze? Perhaps it would be enough to annotate reviews for every new app category and assume that the apps of the same category share enough common features for the model to be generalizable? In case there are annotated reviews available from other domains (reviews of products and services), if and how much do these reviews help to improve the accuracy of aspect extraction from app reviews? These questions are summarized in the second research question: \\

\noindent
\textbf{RQ2:} To what extent is automatic feature extraction from app reviews sensitive to the size and scope of the annotated datasets used? \\

\noindent
To investigate \emph{RQ2} we compare the performance of several training and evaluation procedures. 
The Cross-Category Validation approach used in \emph{RQ1} assumes that the features of the apps belonging to different categories are similar enough and thus the model to extract app features from one category can be trained on the labeled app reviews belonging into different categories. Posing this assumption has two advantages: 
\begin{enumerate*}[label=\arabic*)]
\item we do not need annotated reviews for feature extraction from reviews of a new app that does not yet have labeled data;
\item the size of the annotated dataset available for model training is larger, since we typically have many app reviews from different app categories available.
\end{enumerate*}
In addition to Cross-Category Validation, we also employ other training and evaluation procedures.
First, we restrict ourselves to reviews from apps in the same category. The reason for doing this is that we hope to achieve better performing models when we only use annotated reviews from functionally similar apps (and the app under investigation itself). The downside of this approach is that we might considerably decrease the size of the available annotated data set.
Secondly we perform Stratified Cross-Validation training and evaluation procedure where both training and test folds contain a similar proportion of app features from every app category. In Stratified Cross-Validation, we have the features of the same app category in both training and test set and additionally, we have annotated reviews from other categories in the training set. Stratified Cross-Validation procedure is thus a mixture of Cross-Category Validation used in \emph{RQ1} and training separate models for each app category.
Our experiments show that having annotated training reviews from the test app is not necessary, although including them into training set helps to improve recall with a cost to precision.

Finally, we extend the scope of available annotated data as compared to the situation in \emph{RQ1}. There has been a lot of research on feature extraction from product/service reviews using standard benchmark datasets available with annotated \textsc{Restaurant} and \textsc{Laptop} reviews \citep{toh2014,liu2015,san2015}. Although there are certainly differences between app reviews and these product/service reviews, there may be also many similarities which might enable us to use those benchmark datasets to improve the feature extraction from app reviews. 
Thus, we also experiment with both Cross-Category Validation and Stratified Cross-Validation procedures by extending the training sets with annotated reviews from the \textsc{Restaurant} and \textsc{Laptop} domains.
We show that while using external datasets helps to improve recall, it happens at the cost of significant drop in precision.

The rest of the article is structured as follows. In Section~\ref{sec:related}, we summarize related literature. In Section~\ref{sec:design_vars} we give detailed descriptions of all design variables used in this study. This is followed by Section~\ref{sec:study_design} in which we present the study design for both research questions. In Section~\ref{sec:results}, we present the results of our experiments and answer the research questions. This is followed by a detailed discussion in Section~\ref{sec:discussion} including a discussion of threats to validity. In Section~\ref{sec:conclusion}, we present conclusions and directions of future research.


%% file: related_work.tex
\section{Related work}
\label{sec:related}

A number of studies have proposed methods for extracting app features from app reviews. \citet{guzman2014} used an unsupervised LDA topic modeling approach \citep{blei2003} for automatic extraction of app features from user reviews of seven apps (three from App Store and four from Play Store). The extracted app features were evaluated against human labeled app features. The study of \citet{gu2015} classifies review sentences into categories, such as feature evaluation, praise, feature requests, bug reports and others, and then extracts app features using 26 manually designed rules only from those sentences that belong to the feature evaluation category. 
\citet{johann2017} proposed a rule-based approach \textsc{SAFE} that uses Part of Speech (POS) and sentence patterns for extracting app features from app descriptions and user reviews. 
\citet{malik2018comparing} uses syntactic relations between the features and opinion words for the extraction of app features from user reviews.

A few studies focus on summarizing reviews that are informative for developers using topic models \citep{guzman2015,chen2014,panichella2015,Luiz:2018}. Recently, supervised machine learning approaches have been used to classify app reviews into functional and non-functional requirements \citep{lu2017,kurtanovic2017}. 

Contrary to the fine-grained extraction of app features, the approach of \citet{vu2015mining} extracts all potential keywords from user reviews and ranks them based on the review rating and occurrence frequency. Their approach also has the ability to cluster or expand the set of extracted keywords based on their semantic similarity.
\citet{keertipati2016} extracted nouns as candidate app features from the app review sentences but they did not perform an evaluation to check whether the extracted features actually represent true app features. Our previous study \citep{shah2016} presented the prototype of a web-based tool to identify competing apps and compare them based on the sentiments expressed about the common set of app features. The tool extracts two-word collocations as candidate features without evaluating the extracted features against true features. 

All of the aforementioned studies related to app feature extraction use different techniques, review datasets and/or annotation guidelines; therefore, the results reported in these studies are not directly comparable to each other. Additonally, without having access to different datasets and annotation guidelines, it is difficult to assess the quality of the annotations used to evaluate the systems.
We were able to obtain the evaluation set of \citet{guzman2014} together with its annotation guidelines and we use this dataset as one of the annotated experimental training sets (\textsc{Guzman} dataset) in our study.

In the literature, a plethora of research \citep{kang2017,qiu2011} is dedicated to automatic extraction of features from product reviews in the \textsc{Laptop} and \textsc{Restaurant} domains. The best results have been achieved using supervised learning approach such as Conditional Random Fields (CRF) and Recurrent Neural Networks (RNN) \citep{pontiki2016,san2015,toh2014,liu2015,poria2016}.
To the best of our knowledge, the only study that has used supervised sequence modeling (CRF) for automatic extraction of app features from app reviews has been performed by \citet{sanger2016} on German app reviews. 

Since supervised machine learning approaches requires annotated datasets for training, the performance of the model is potentially largely impacted by the quality of annotations.
Thus, we also adopt a supervised sequence learning (CRF) approach in this work to evaluate the quality of annotation guidelines by assessing the changes in model performance brought about by the changes in annotation guidelines.
The German dataset and annotation guidelines of \citet{sanger2016} are publicly available and these were used in this study in comparison with \textsc{Guzman} dataset and annotation guidelines. 
We also use the annotation guidelines of \citet{sanger2016} to create a new dataset with manually labeled app features in user reviews.


%% file: design_variables.tex
\section{Design variables}
\label{sec:design_vars}

We study both Research Questions by evaluating the results of several empirical experiments on app review data human-annotated with app features. These experiments include several variables, some of which can be considered as given and others are manipulated to study the Research Questions. Table~\ref{tab:variables} summarizes all variables. Below we describe all of them in more detail.

\begin{table}
\centering
\begin{tabularx}{\linewidth}{X}
\toprule
\multicolumn{1}c{\bf GIVEN DESIGN VARIABLES} \\
\midrule
\bf 1. Annotation guidelines \\
\begin{enumerate}[label=\alph*)]
\item \textsc{Sanger} annotation guidelines (German and English)
\item \textsc{Guzman} annotation guidelines (English)
\end{enumerate} \\
\cmidrule[\lightrulewidth](lr){1-1}
\bf 2. Annotated datasets \\
\begin{enumerate}[label=\arabic*)]
\item \textsc{Sanger} dataset \citep{sanger2016}, annotated using \textsc{Sanger} guidelines
\item \textsc{Shah} dataset: annotated by two annotators using translated \textsc{Sanger} guidelines.
\begin{enumerate}[label=\alph*)]
\item \textsc{Shah-I}: \textsc{Shah} dataset labeled by annotator I.
\item \textsc{Shah-II}: \textsc{Shah} dataset labeled by annotator II.
\end{enumerate} 
\item \textsc{Guzman} dataset \citep{guzman2014}, annotated using \textsc{Guzman} guidelines
\end{enumerate} \\
\cmidrule[\lightrulewidth](lr){1-1}
\bf 3. Modeling approach \\
\vspace{0.5mm}
CRF model with the following features extracted from the current word and its context of two preceding and two following words: \\
\begin{enumerate}[label=\alph*)]
\item the words themselves
\item POS of the words in the sentence
\item one to four character prefixes and suffixes of the words
\item the position of the words
\item the stylistics of each word (e.g. case, digit, symbol, alphanumeric) 
\end{enumerate} \\
\cmidrule[\lightrulewidth](lr){1-1}
\bf 4. Evaluation methods \\
\begin{enumerate}[label=\alph*)]
\item Token-based exact match
\item Token-based partial match
\item Type-based exact match
\item Type-based partial match
\end{enumerate} \\
\midrule
\multicolumn{1}c{\bf MANIPULATED DESIGN VARIABLES} \\
\midrule
\bf 5. Data processing flow \\
\begin{enumerate}[label=\textbf{Step \arabic*},leftmargin=17mm]
\item (Pre-processing): remove non-consecutive app features, remove app reviews with no annotated features
\item (Simulation step I): Remove \emph{pseudo}-features
\item (Simulation step II): Remove app features that do not contain a NOUN
\item (Simulation step III): Remove app features that are longer than three words
\end{enumerate} \\
\cmidrule[\lightrulewidth](lr){1-1}
\bf 6. Training procedures \\
\begin{enumerate}[label=\alph*)]
\item \textsc{CCV:} \textbf{C}ross-\textbf{C}ategory \textbf{V}alidation on the full dataset
\item \textsc{appCat:} 10-fold cross-validation over single \textbf{app Cat}egory 
\item \textsc{SCV:} \textbf{S}tratified 10-fold \textbf{C}ross-\textbf{V}alidation on the full dataset 
\item \textsc{CCV-Ext:} \textbf{C}ross-\textbf{C}ategory \textbf{V}alidation with \textbf{Ext}ernal \textsc{Laptop} and \textsc{Restaurant} datasets
\item \textsc{SCV-Ext:} \textbf{S}tratified 10-fold \textbf{C}ross-\textbf{V}alidation on the full dataset with \textbf{Ext}ernal \textsc{Laptop} and \textsc{Restaurant} datasets
\end{enumerate}\\

\bottomrule
\end{tabularx}
\caption{The summary of design variables used in the experiments.}
\label{tab:variables}
\end{table}

\subsection{Annotation guidelines}
\label{sec:annotation_guidelines}

Our experimental datasets (described more thoroughly in the next subsection) are annotated using two distinct set of annotation guidelines: \textsc{Guzman} Annotation Guidelines and \textsc{Sanger} Annotation Guidelines.

\textsc{Guzman} Annotation Guidelines\footnote{\url{https://mast.informatik.uni-hamburg.de/wp-content/uploads/2014/03/Coding_Guide.pdf}} were developed by \citet{guzman2014} to annotate their evaluation set. These annotation guidelines define an app feature as a description of specific app functionality visible to the user (such as \emph{uploading files} or \emph{sending emails}), a specific screen of the app, a general quality of the app (such as \emph{time needed to load} or \emph{size of storage}) or a specific technical characteristic (e.g. a network protocol or HTML5). The annotation guidelines encourage to annotate the exact words used in the text but do not explicitly demand it. 
The guidelines also explicitly allow the annotation of app features consisting of non-consecutive words.

\textsc{Sanger} Annotation Guidelines\footnote{available from \url{http://www.romanklinger.de/scare/}} were developed by \citet{sanger2016} to annotate both app features, subjective phrases and relationships between them. We translated these guidelines from German to English using Google Translate. \textsc{Sanger} annotation guidelines define as an app feature \quotes{anything that is part of the application or in some form connected with the app}. This includes existing and requested app features, bugs and errors as well as entities referring to non-functional features such as usability, design, price, license, permissions, advertisements and updates. These guidelines explicitly instruct to annotate the mentions of the app itself as a feature. Instructions also ask to annotated implicit features represented by a single verb such as \emph{runs}. Annotators are encouraged to keep the annotated features as short as possible although a particular length limit is not set. The \textsc{Sanger} guidelines specifically require not to include function words into annotated app features, which probably also influences the length of the annotated app features. Although no explicit mention about annotating consecutive vs non-consecutive words as features is made, all example features only consist of consecutive words.

Although both annotation guidelines can be used to label the same information---features in app reviews---they have several crucial differences which can influence how well the data annotated with these guidelines can be used to train a model for automatic aspect extraction.
\begin{enumerate}[1),noitemsep]
\item Using feature annotations not comprised of exact words used in the review text will make any automatic use of these annotations very difficult. Although this practice is discouraged in \textsc{Guzman} guidelines, it is not directly prohibited.
\item Annotating non-consecutive app features, allowed in \textsc{Guzman} guidelines, restricts the types of models that can be used. In particular, sequence tagging models, such as Conditional Random Fields and Recurrent Neural Networks which produce state-of-the-art results on the feature extraction task \citep{liu2015}, can only process aspects that consist of consecutive words.
\item The instruction in \textsc{Sanger} guidelines to annotate mentions and references to the app itself is most probably motivated by the particular task of \citet{sanger2016} to learn to extract both app features and their relations to subjective phrases. In the context of plain app feature extraction these features can be considered \emph{pseudo}-features, as they not give any useful information to the app developers. 
\item Similarly, the instruction in \textsc{Sanger} guidelines to annotate standalone abstract verbs such as \emph{runs} is also probably motivated by the joint task of learning both app features and subjective phrases. In the context of app feature extraction these aspects will likely cause problems because they are difficult to distinguish from other generic verbs not labeled as app features. Also, these very generic app features are likely of very little value to the developers.
\end{enumerate}

\subsection{Annotated datasets}
\label{sec:annotated_datasets}

We have at our disposal four annotated app review datasets: \textsc{Guzman} dataset, \textsc{Sanger} dataset and two versions of \textsc{Shah} datasets. In Table~\ref{tab:init_data} we present some characteristics for each of these datasets. Note that we do not show data per individual app but aggregated per app category. Each review dataset is characterized using the following information: 
\begin{enumerate}[label=\alph*),noitemsep]
\item the total number of reviews;
\item the total number of sentences in all reviews;
\item the total number of annotated app features in (tokens);
\item the number of distinct app features (types);
\item the number of app features consisting of a single word only; 
\item the number of app features consisting of at least two words; and
\item the type-token ratio of annotated app features (the number of feature types divided by the number of feature tokens).
\end{enumerate}

The \textsc{Guzman} dataset\footnote{The dataset was obtained from the authors of \citet{guzman2014}} was used as an evaluation set in the study performed by \citet{guzman2014}. It contains annotated app reviews in English language from six different app categories.
Most app categories contain reviews from one app only: AngryBirds from Games category, TripAdvisor from Travel category, PicsArt from Photography category, Pinterest from Social category and Whatsapp from Communication category. The only exception is the Productivity category which contains reviews from two apps: Dropbox and Evernote. According to \citet{guzman2014}, 400 reviews were annotated for each app. However, from Table~\ref{tab:init_data} it can be seen that the \textsc{Guzman} dataset includes less than 400 reviews in each category. This is because the \textsc{Guzman} dataset only contains reviews with at least one annotated app feature.\footnote{From the dataset we obtained from the authors of \citet{guzman2014} the reviews without any annotated app features had been filtered out.}

The \textsc{Sanger} dataset\footnote{also available from \url{http://www.romanklinger.de/scare/}} was used in the study performed by \citet{sanger2016}. It contains reviews in German language. The \textsc{Sanger} dataset contains the same number of reviews in each app category. The reviews in each category come from 10-15 different apps but the origin of each particular review is unknown to us. In addition to app features, this dataset is also annotated with subjective phrases and relations between features and subjective phrases. However, in our study we only use the annotated app features and ignore the other annotations.

In addition to the two datasets available from other researchers, we created the \textsc{Shah} dataset. The app reviews included in this dataset were selected by randomly sampling 500 reviews per app from the total pool of reviews assembled by \citet{guzman2014} for their study. The app categories and apps in each category are the same as in the \textsc{Guzman} dataset. Similar to \citet{guzman2014}, since each app has its own user rating distribution, we used a stratified sampling procedure to sample the reviews using the distribution over ratings as stratum.
For instance, assume that the user rating distribution over the DropBox app reviews in our dataset is as follows: 
30\% of the reviews have 5 stars, 40\% of the reviews have 4 stars, 20\% of the reviews have 3 stars, 8\% of the reviews have 2 star, and 2\% of the reviews have 1 star. In this case, the stratified random sampling of 100 reviews from a pool of 1000 reviews for the DropBox app randomly selects 30 reviews from 5-star reviews, 40 from 4-star reviews, 20 from 3-star reviews, 8 from 2-star reviews, and 2 from 1-star reviews. All reviews from the \textsc{Shah} dataset were independently annotated by two annotators\footnote{Both annotators were software engineering bachelor students at the University of Tartu} according to the \textsc{Sanger} guidelines with both app features and subjective phrases, although for this study only the app feature annotations are used. 
For measuring the inter-annotator agreement, we adopted the Dice coefficient \citep{pavlopoulos2014}, which ranges between 0 and 1 where 1 means total agreement and 0 total disagreement. The Dice coefficient value between the two annotators was 0.28 which denotes a low agreement between the annotators.
Because of that, we decided to treat the annotations of both annotators as different datasets. Thus, we have two annotated \textsc{Shah} datasets: \textsc{Shah-I} and \textsc{Shah-II} containing the annotations of the first and the second annotator respectively.

\begin{table}[!htbp]
\small
\centering
\begin{tabularx}{\textwidth}{lcrrrrrrr}
\toprule
\multirow{3}{*}{\bf Category} & \multicolumn{1}{c}{\bf Number} & \multicolumn{1}{c}{\bf Number} & \multicolumn{1}{c}{\bf App}  & \multicolumn{1}{c}{\bf App} & \multicolumn{1}{c}{\bf Single}  & \multicolumn{1}{c}{\bf Multi}  & \multicolumn{1}{c}{\bf Type-} & \multicolumn{1}{c}{\bf Features}
\\
 & \multicolumn{1}{c}{\bf of} & \multicolumn{1}{c}{\bf of} & \multicolumn{1}{c}{\bf feature} & \multicolumn{1}{c}{\bf feature} & \multicolumn{1}{c}{\bf word} & \multicolumn{1}{c}{\bf word} & \multicolumn{1}{c}{\bf token} & \multicolumn{1}{c}{\bf per}
 \\
 & \multicolumn{1}{c}{\bf reviews} & \multicolumn{1}{c}{\bf sents} & \multicolumn{1}{c}{\bf tokens} & \multicolumn{1}{c}{\bf types} & \multicolumn{1}{c}{\bf feats} & \multicolumn{1}{c}{\bf feats} & \multicolumn{1}{c}{\bf ratio} & \multicolumn{1}{c}{\bf review}\\
\midrule
\\
\multicolumn{2}{l}{a) \textsc{Guzman} dataset (English)} \\
\cmidrule[\lightrulewidth](lr){1-9}
Games & 238 & 718 & 381 & 254 & 112 & 269 & 0.67 & 1.60\\
Productivity& 602 & 2069 & 1188 & 932 & 328 & 860 & 0.78 & 1.97\\
Travel & 265 & 720 & 518 & 379 & 169 & 349 & 0.73 & 1.95\\
Photography & 138 & 289 & 184 & 127 & 83 & 101 & 0.75 & 1.33\\
Social & 146 & 379 & 257 & 192 & 98 & 159 & 0.75 & 1.76\\
Communication & 90 & 192 & 137 & 114 & 60 & 77 & 0.83 & 1.52\\
\cmidrule[\lightrulewidth](lr){1-9}
\multicolumn{1}{r}{\bf Total} & \bf 1479 & \bf 4367 & \bf 2665 & \bf 1998 & \bf 850 & \bf 1815 & \bf 0.75 & \bf 1.80\\
\midrule
\\
\multicolumn{2}{l}{b) \textsc{Shah-I} dataset (English)} \\
\cmidrule[\lightrulewidth](lr){1-9}
Games & 500 & 901 & 283 & 84 & 167 & 116 & 0.30 & 0.57\\
Productivity & 1000 & 2159 & 480 & 198 & 295 & 185 & 0.41 & 0.48\\
Travel & 500 & 866 & 266 & 93 & 185 & 81 & 0.35 & 0.53\\
Photography & 500 & 680 & 266 & 61 & 212 & 54 & 0.23 & 0.53\\ 
Social & 500 & 790 & 286 & 66 & 242 & 44 & 0.23 & 0.57\\
Communication & 500 & 574 & 159 & 34 & 133 & 26 & 0.21 & 0.32\\
\cmidrule[\lightrulewidth](lr){1-9}
\multicolumn{1}{r}{\bf Total} & \bf 3500 & \bf 5970 & \bf 1740 & \bf 536 & \bf 1234 & \bf 506 & \bf 0.31 & \bf 0.50 \\
\midrule
\\
\multicolumn{2}{l}{c) \textsc{Shah-II} dataset (English)} \\
\cmidrule[\lightrulewidth](lr){1-9}
Games & 500 & 901 & 140 & 79 & 48 & 92 & 0.56 & 0.28 \\
Productivity & 1000 & 2159 & 469 & 295 & 217 & 252 & 0.63 & 0.47 \\
Travel & 500 & 866 & 185 & 119 & 89 & 96 & 0.64 & 0.37 \\
Photography & 500 & 680 & 81 & 53 & 46 & 35 & 0.65 & 0.16\\
Social & 500 & 790 & 144 & 75 & 93 & 51 & 0.58 & 0.13\\
Communication & 500 & 574 & 65 & 38 & 43 & 22 & 0.61 & 0.31\\
\cmidrule[\lightrulewidth](lr){1-9}
\multicolumn{1}{r}{\bf Total} & \bf 3500 & \bf 5970 & \bf 1084 & \bf 659 & \bf 536 & \bf 548 & \bf 0.61 & \bf 0.31\\
\midrule
\\
\multicolumn{2}{l}{d) \textsc{Sanger} dataset (German)} \\
\cmidrule[\lightrulewidth](lr){1-9}
Alarm Clocks & 160 & 349 & 203 & 92 & 176 & 27 & 0.45 & 1.27 \\
Fitness Tracker & 160 & 270 & 194 & 107 & 153 & 41 & 0.55 & 1.21 \\
Games & 160 & 253 & 162 & 77 & 144 & 18 & 0.48 & 1.01 \\
Instant Messengers &  160 & 247 & 156 & 90 & 125 & 31 & 0.58 & 0.97 \\
Music Player & 160 & 304 & 213 & 108 & 179 & 34 & 0.51 & 1.33 \\
Navigation & 160 & 331 & 181 & 103 & 146 & 35 & 0.57 & 1.13 \\
News & 160 & 160 & 234 & 115 & 202 & 32 & 0.49 & 1.46 \\
Office tools & 160 & 348 & 219 & 132 & 171 & 48 &  0.60 & 1.37 \\
Social Networks & 160 & 237 & 119 & 59 & 107 & 12 & 0.50 & 0.74 \\
Sport News & 160 & 282 & 179 & 97 & 147 & 32 & 0.54 & 1.12 \\
Weather & 160 & 263 & 181 & 88 & 159 & 22 & 0.49 & 1.13 \\
\cmidrule[\lightrulewidth](lr){1-9}
\multicolumn{1}{r}{\bf Total} & \bf 1760 & \bf 3044 & \bf 2041 & \bf 1068 & \bf 1709 & \bf 332 & \bf 0.52 & \bf 1.16 \\
\bottomrule
\end{tabularx}
\caption{Characteristics of the annotated app review datasets.}
\label{tab:init_data}
\end{table}

Based on the statistics presented in Table~\ref{tab:init_data} one can see several differences between the datasets. First of all, the \textsc{Guzman} dataset is quite different from others when comparing the number of single-word features and multi-word features with multi-word features occurring twice as often as single-word features. Conversely, both \textsc{Shah-II} and \textsc{Sanger} datasets have more single-word features than multi-word features while in \textsc{Shah-I} dataset these numbers are actually balanced. There can be several reasons for the difference in these numbers, including the manner how each particular annotator interpreted the annotation guidelines given to them, but we believe that this difference might also characterize the differences in the annotation guidelines themselves.

The second main difference manifests itself in the average number of app features per review (the last column in Table~\ref{tab:init_data}). This quantity is the largest for the \textsc{Guzman} dataset and the smallest for the \textsc{Shah} datasets with \textsc{Sanger} dataset falling in between. We attribute this difference to fact that the \textsc{Guzman} dataset, as explained above, only consists of reviews that contain at least one annotated app feature while the other datasets may also contain reviews without any annotated app features.

Thirdly, also the type-token ratio of app features the largest on the \textsc{Guzman} dataset, which shows that the proportion of distinct app features is largest on this dataset. This can be explained by the large number of multi-word app features because the longer the features the more likely they consist of a unique sequence of words.

\subsection{Modeling approach}
\label{sec:modeling_approach}

We adopt the Conditional Random Field (CRF) \citet{lafferty2001} supervised learning method to train the models for all our experiments. CRF is a sequence tagging model which tags each word in the app review text with a label. Similarly to \citet{sanger2016} we use the BIO labeling scheme, where the tag \textbf{B} is used to annotate the first word of each app feature, \textbf{I} labels the rest of the words \emph{inside} the app feature and the label \textbf{O} is used to tag all words that are \emph{outside} of the app feature. We use an implementation based on CRFSuite\footnote{\url{https://github.com/pdsujnow/opinion-target}} which was used as a CRF baseline by \citet{liu2015} on \textsc{Laptop} and \textsc{Restaurant} product review datasets. 

The hand-crafted features features used in the CRF model are the same as used by \citet{liu2015}, they are summarized in Table~\ref{tab:variables}.
The study by \citet{liu2015} showed that using word embeddings---low-dimensional distributed vector representations of words \citep{mikolov2013}---as additional features in the CRF model improves model performance. Therefore, we included word embeddings as features in all the experiments. For the datasets in English language (\textsc{Guzman} and \textsc{Shah} datasets), we used SENNA embeddings \citet{collobert2011}.\footnote{https://ronan.collobert.com/senna/} For the \textsc{Sanger} dataset in German language, we used the embeddings\footnote{\url{https://spinningbytes.com/resources/word-embeddings/}} trained on Wikipedia articles. 

\subsection{Evaluation procedures}
\label{sec:eval_procedures}
We evaluate all results by computing the precision, recall and F1-score of the predicted app features. Because the app feature annotations themselves can often be noisy and and ambiguous, which manifests itself in low agreement between annotators (for instance \citep{guzman2014} reported an agreement of 53\% on annotated app features), we adopt both token and type-based evaluation methods using both exact and partial matching between predicted and human-annotated features (see Part 4 in Table~\ref{tab:variables}).

\paragraph{Exact match} requires the predicted and annotated app features to match exactly. For instance, if the annotated feature is \emph{to upload video} then in order to count a match the predicted app feature must consist of exactly the same words. If the model predicts \emph{upload video} as feature leaving the particle \emph{to} untagged the prediction is counted as false positive under the exact match scheme.

\paragraph{Partial match} allows some mismatch when comparing the predicted app features with the human-annotated features in the evaluation set. A difference of one word is allowed. Under the partial match scheme, the predicted feature \emph{upload video} will be counted as true positive even when the human-annotated feature is \emph{to upload video}, whereas the predicted feature \emph{video} would be counted as false positive because it differs from the human-annotated feature in more than one word. Similarly, a predicted feature \emph{failed to upload video} would be counted as true positive under partial match but an even longer predicted feature such as \emph{failed to upload video to} will be counted as false positive.

\paragraph{Token-based evaluation} counts and evaluates each instance of an app feature separately. This approach enables to distinguish between features and non-features expressed with the same sequence of words. For instance, if an app feature \emph{upload video} occurs several times in different reviews then each instance of that feature will be counted separately. Moreover, it might also happen that depending on context, not all \emph{upload video} word bigrams are annotated as app features. Token-based evaluation enables to penalize the wrongly predicted non-features and foster 
the correctly predicted app features in each instance separately.

\paragraph{Type-based evaluation} counts and evaluates each app feature type only once, regardless of how many times it occurs in the review texts. In order to cluster together different instances of the same app feature type, the features are first lemmatized using \textit{Snowball}\footnote{http://www.nltk.org/\_modules/nltk/stem/snowball.html} stemmer available in NLTK library and then matched based on their lemmas. The type-based evaluation procedure is unbiased by the frequencies of the single app feature types. While the token-based evaluation measures can become artificially high when the annotated training and test set contain a single high-frequency simple one-word app feature, the type-based evaluation gives equal credit to all different app features, regardless of their frequency.

\subsection{Data processing}
\label{sec:data_processing}

The data processing steps adopted in experiments include one preprocessing step and three steps for simulating the changes in annotation guidelines. 

\paragraph{The Preprocessing Step} is necessary to unify all experimental annotated datasets to bring them to similar starting point. First of all, \textsc{Guzman} dataset can also include annotations of non-consecutive app features. Because the CRF model can only learn app features consisting of consecutive words, we remove all non-consecutive app features from \textsc{Guzman} dataset because leaving them in would put the \textsc{Guzman} datasets and \textsc{Guzman} annotation guidelines into a disadvantaged position compared to the \textsc{Sanger} and \textsc{Shah} datasets annotated with \textsc{Sanger} guidelines where annotated app features always consist of only consecutive words.
After that, we remove all reviews from the datasets that do not contain any annotated app feature. We do this because the annotated \textsc{Guzman} dataset we obtained from the authors of \citep{guzman2014} is a subset of all the reviews originally annotated by \citet{guzman2014}---the annotated reviews containing no app features have been left out. Although removing such reviews biases the datasets doing so makes the \textsc{Sanger} and \textsc{Shah} datasets comparable to the \textsc{Guzman} dataset in terms of app feature distribution over reviews.

\paragraph{The Simulation Step I} removes all app features that refer to app itself either by the app name or by explicitly using the words such as \emph{app} or \emph{application} and other similar pseudo-features. This step simulates the change in the \textsc{Sanger} annotation guidelines such that the command to annotate the references to the app itself are removed. Because \textsc{Guzman} annotation guidelines do not require to annotate such pseudo-features this simulation step only changes the annotations of the \textsc{Sanger} and \textsc{Shah} datasets. After this step, the reviews without any annotated app features are removed again to ensure that the feature distributions over all datasets are similar.

\paragraph{The Simulation Step II} removes all app features that do not contain a noun. The rationale behind this simulation step is that useful app features should be specific enough and this can only be achieved by requiring the presence of a noun phrase. For instance, an app feature such as \emph{to upload}, which consists of a particle and a verb and does not include a noun, is too non-specific to understand what kind to functionality the feature refers to. Thus, after this simulation step, these kinds of word sequences are not considered as app features anymore, whereas a similar word sequence \emph{to upload video}, which specifies the action with a noun, will be kept. This simulation step mostly removes short generic app features annotated according to \textsc{Sanger} guidelines.

\paragraph{The Simulation Step III} removes all app features that are longer than three words. We believe that useful features cannot be too long because otherwise they become too specific and noisy. We attempt to simulate the change in annotation guidelines that would limit the maximum length of an app feature to three words with a very crude heuristic that just removes the longer features from the dataset. Although a better heuristic would be to develop a set of rules to shorten the app features appropriately we opted here for the simplest strategy, believing that it will be good enough for our purpose of testing the potential effect of such a guideline.

\subsection{Training procedures}
\label{sec:training_procedures}

We use three different training procedures to study our Research Questions: Cross-category validation, 10-fold cross-validation over single apps, and cross-category validation including additional external training data. 
 
\paragraph{Cross-category validation (\textsc{CCV}):} The \textsc{CCV} training procedure assumes that the annotated training data consists of app reviews belonging to several different app categories. Then the reviews of each app category are held out in turn and the model is trained on the reviews of the rest of the app categories. Finally the trained model is evaluated on the held-out reviews. This training regime assumes that the reviews of different app categories share enough common information that a model trained on the reviews belonging to one set of apps or app categories will generalize to the apps or app categories whose reviews the model has not seen during training. We use cross-category validation instead of cross-app validation because both in \textsc{Guzman} and \textsc{Shah} datasets, with one exception we have the reviews of just one app in each of the app categories. In \textsc{Sanger} dataset, although each category contains reviews from several apps, we do not have the app name annotations attached to each review and thus we could not separate the reviews of different apps into different subsets.

\paragraph{10-fold cross-validation over single app category (\textsc{appCat}):} 
In case the different app categories do not share enough common app features, the model trained using \textsc{CCV} would not be able to generalize to the reviews of a new app category. Therefore, the training procedure \textsc{appCat} is designed with an expectation that the annotated reviews of a particular app category are necessary to learn a model that is able to extract app features from new reviews of the same app category. The \textsc{appCat} procedure treats the reviews of each app category as distinct datasets and performs a 10-fold cross-validation over each of those data sets. In 10-fold cross-validation, the reviews of an app category are equally partitioned into ten random samples of equal size. One sample is held out for testing, the model is trained on the rest of the nine samples and evaluated on the held-out sample. This procedure is repeated ten times until all samples are held out in turn and then the obtained results are averaged. 

The advantage of \textsc{appCat} compared to \textsc{CCV} is the fact that unlike \textsc{CCV}, which always evaluates the models on the reviews of a different set of apps than those used to train the model, \textsc{appCat} trains and evaluates the model on the reviews of the same app category. This way there are more chances that the model has seen the same or similar app features during training that it encounters during testing.
The main critique of this procedure is that the training data is now very small, consisting of the reviews of a single app only, and when comparing these results with \textsc{CCV}, it is highly likely that any differences are due to the difference in training set sizes, rather than the fact that now training and testing sets contain the reviews of the same app. 

\paragraph{Stratified 10-fold cross-validation (\textsc{SCV}):} 
One possible option to improve the \textsc{appCat} is to use stratified cross-validation on the whole training set instead, using the app category as stratum. This way the sizes of the training folds would be similar to those used in the \textsc{CCV} setting and at the same time, the reviews of all apps would occur both in training and test folds. Naturally, the downside is that annotations in all app categories are required for applying the \textsc{SCV} procedure.

\paragraph{\textsc{CCV} with external \textsc{Laptop} and \textsc{Restaurant} datasets (\textsc{CCV-Ext}):}
Supervised machine learning approaches are expected to give better results when supplied more training data. Because the amount of annotated app reviews is limited, we are interested in whether using additional annotated training data from external domains would still be helpful. The \textsc{CCV-Ext} procedure is otherwise identical to the \textsc{CCV} procedure described above with a difference that for training each model the external annotated datasets\footnote{\url{http://alt.qcri.org/semeval2014/task4/index.php?id=data-and-tools}} of \textsc{Laptop} and \textsc{Restaurant} product reviews are included in the training set. Although product reviews and app reviews are very different we hope there are some similarities that might help in improving the model generalization capabilities.

\paragraph{SCV with external \textsc{Laptop} and \textsc{Restaurant} datasets (\textsc{SCV-Ext}):}
To check whether supplying more training data further improves the \textsc{SCV} procedure, we include \textsc{Laptop} and \textsc{Restaurant} product reviews in the training set. This is similar to what we do with the \textsc{CCV} procedure.

%% file: study_design.tex
\section{Study design}
\label{sec:study_design}


For studying the Research Question \emph{RQ1} we use all Given Design Variables 1-4 and adopt \textsc{CCV} as training procedure to explore the effects of the Data Processing steps simulating the changes in the annotation guidelines. We rely on an assumption that there is a correlation between the quality of the training data annotations and the accuracy of the app feature extraction model. Thus, we use the model accuracy on the test set to assess the usefulness of the proposed changes in the annotation guidelines.
In particular, we train and evaluate CRF-based app feature extraction models on all our annotated datasets after each Data Processing steps and assess their accuracies to approximate how each step affects the quality of the annotations. 

For studying the Research Question \emph{RQ2}, we start with the simulated annotations obtained as a result of \emph{RQ1}, i.e. all datasets have been processed with all Data Processing steps. We employ again the Given Design Variables 2-4 and now vary the training procedures (5a-5c) to assess how the amount and scope of training data affects the app feature extraction performance. Table~\ref{tab:experiments} summarizes the design variables used to study both Research Questions.

\begin{table}[h]
\small
\centering
\begin{tabularx}{\linewidth}{XXX}
\toprule
\multicolumn{1}{c}{\bf Design Variable} & \multicolumn{1}{c}{\bf RQ1} & \multicolumn{1}{c}{\bf RQ2} \\
\cmidrule[\lightrulewidth](lr){1-3}
\bf 1. Annotation Guidelines & \textsc{Guzman} and \textsc{Sanger} Annotation Guidelines & Simulated annotations obtained as a result of the Simulation Step III \\
\cmidrule[\lightrulewidth](lr){1-3}
\bf 2. Initial datasets & All datasets & All datasets\\
\cmidrule[\lightrulewidth](lr){1-3}
\bf 3. Modeling approach & CRF & CRF \\
\cmidrule[\lightrulewidth](lr){1-3}
\bf 4. Evaluation procedures &
\begin{itemize}
\item Token-based exact match
\item Type-based exact match
\item Token-based partial match
\item Type-based partial match
\end{itemize} 
 &
 \begin{itemize}
\item Token-based exact match
\item Type-based exact match
\item Token-based partial match
\item Type-based partial match
\end{itemize} \\
\cmidrule[\lightrulewidth](lr){1-3}
\bf 5. Data Processing & \begin{itemize}
\item Preprocessing Step
\item Simulation Step I
\item Simulation Step II
\item Simulation Step III
\end{itemize}& \\
\cmidrule[\lightrulewidth](lr){1-3}
\bf 6. Training procedures & \begin{itemize}
\item \textsc{CCV}
\end{itemize} & 
\begin{itemize}
\item \textsc{CCV}
\item \textsc{appCat}
\item \textsc{SCV}
\item \textsc{CCV-Ext}
\item \textsc{SCV-Ext}
\end{itemize}\\
\bottomrule
\end{tabularx}
\caption{Summary of the experimental design used to study both Research Questions.}
\label{tab:experiments}
\end{table}



%% file: results.tex
\section{Results}
\label{sec:results}

In this section, we present the results of our study (as described in Sections~\ref{sec:design_vars} and ~\ref{sec:study_design}) and answer the research questions. For better readability, we only show aggregated results for each dataset. 
The detailed results of our study at app category level can be found in Appendix A (RQ1), Appendix B (RQ2) and Appendix C (RQ2), respectively.

\subsection{RQ1: How sensitive is automatic feature extraction from app reviews to the used annotation
guidelines?}

\begin{figure}[h!]
\centering
\includegraphics[width=0.90\linewidth]{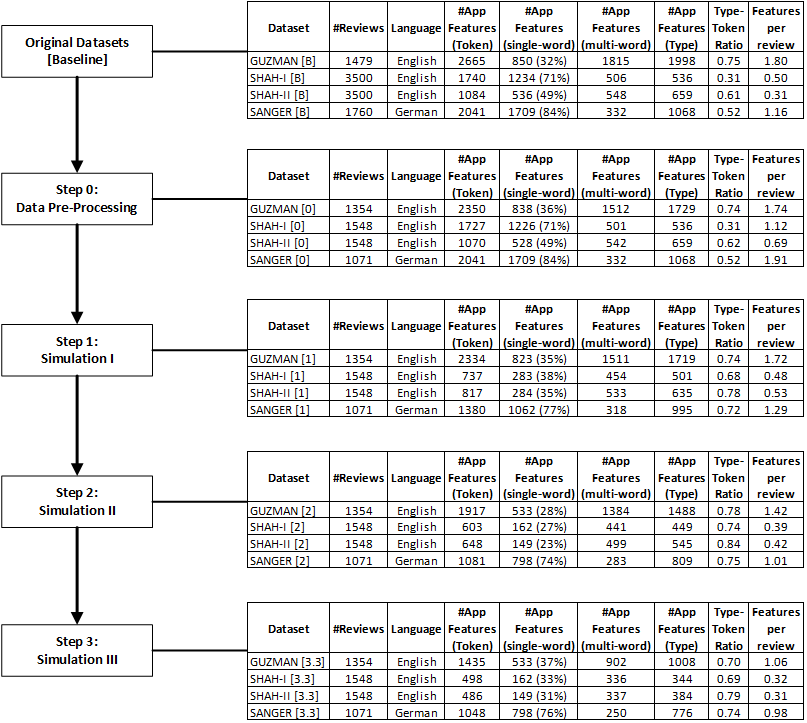}
\caption{Series of modifications performed on the labeled datasets}
\label{fig:figure1}
\end{figure}

Figure~\ref{fig:figure1} shows how the characteristics of the labeled datasets used in our study change step-by-step from the starting situation (Baseline) to Step Simulation III-3, i.e., after all data processing steps have been performed and the best performance has been achieved with training procedure \textsc{CCV} (Cross-Category Validation). The details about the data processing steps as well as the training procedure \textsc{CCV} have been described in Sections~\ref{sec:data_processing} and ~\ref{sec:training_procedures}, respectively.

The annotated datasets at the Baseline correspond to those described in Section~\ref{sec:annotated_datasets}. Several phenomena can be observed when comparing evolution of the datasets' characteristics from Baseline to Step 3. Due to the nature of the data processing steps, the number of app features steadily decrease in all datasets both token-wise and type-wise. Also, several of the characteristics of the four datasets converge. For example, at the Baseline, the type-token ratio of app features varies in the range [0.31, 0.75], while after Step 3, the variation is reduced to the range [0.69, 0.79]. In other words, in Step 3, most of the feature instances occur only once or twice in each of the datasets.
Similarly, the average number of features per review, which initially varies in the range [0.31, 1.80], reduces to a range of [0.31, 1.06] after the Step 3.
For the \textsc{Guzman}, \textsc{Shah-I}, and \textsc{Shah-II} datasets the portion of single-word features converges from variation in range [0.32, 0.71] to variation in range [0.31, 0.37]. Only for the \textsc{Sanger} dataset, the portion of single-word features stays high (with a small reduction from 0.84 to 0.76). One explanation for the high portion of single-word features in the \textsc{Sanger} dataset could be that the German language allows for noun compositions replacing multiple-word noun phrases. 

In the following, we present the performance corresponding to Steps 0, 1, 2, and 3 in Table~\ref{tab:rq1_res}. For each experiment, the table shows the precision, recall and F1-measure of our models when applying the four different evaluation procedures \textsc{Exact match (Token)}, \textsc{Partial match (Token)}, \textsc{Exact match (Type)}, and \textsc{Partial match (Type)} as described in Section~\ref{sec:eval_procedures}.

The first row in each section of the Table~\ref{tab:rq1_res} (Pre-processing) shows the performance after filtering out non-consecutive app features and removing reviews that do not mention app features.
On all datasets precision is consistently better than recall for all evaluation procedures, and partial matching, as expected, yields better performance than exact matching for both token- and type-based evaluation. In particular for token-based evaluation, performance varies largely between datasets. The models built using the \textsc{Guzman} and \textsc{Shah-II} datasets clearly perform worse than those built using the \textsc{Shah-I} and \textsc{Sanger} datasets. 
When looking at the dataset characteristics, one sees the following similarities between the datasets on which the models perform better as compared to the datasets where the models perform worse:
\begin{itemize}
\item The \textsc{Shah-I} and \textsc{Sanger} datasets have a larger share of single-word app features (71\% and 84\%) than the \textsc{Guzman} and \textsc{Shah-II} datasets (36\% and 49\%).
\item The \textsc{Shah-I} and \textsc{Sanger} datasets have a lower type-token ratio (0.31 and 0.52) than the \textsc{Guzman} and \textsc{Shah-II} datasets (0.74 and 0.62). 
\end{itemize}

\indent After Step 0, it seems that the language used in the review datasets (German in the case of the \textsc{Sanger} dataset and English in the case of the \textsc{Shah-I} dataset) does not have a distinguishing impact on model performance. The impact of the two annotation guidelines seems to be mixed. 
Even though we used the translated \textsc{Sanger} Annotation Guidelines when annotating the \textsc{Shah} dataset, the performances of \textsc{Shah-I}-based and \textsc{Shah-II}-based models are very different. The performance of the \textsc{Shah-II}-based model is even worse than the performance of the \textsc{Guzman}-based model where a different annotation guideline was used. 
We speculated that one possible reason for the difference in performance could be that the \textsc{Sanger} Annotation Guidelines explicitly instruct annotating references to the app itself as a feature. Following this instruction automatically increases the number of single-word features and lowers the type-token ratio as the repeated mentioning of the app itself increases the token count but not the type count. 
When inspecting the \textsc{Shah-II} dataset, we noticed that the annotator seemed to have ignored this instruction. Since the frequent annotation of references to the app itself in a review seems to artificially boost the performance of the feature extraction models, while it does not have any practical value to correctly predict the occurrence of a feature referring to the app itself, we decided to remove the annotations of app-references in our datasets.

\begin{table}[]
\centering
\begin{tabular}{l@{\hspace{0.5\tabcolsep}}rrrrrrrrrrrr}
\toprule
\bf Processing & \multicolumn{3}{c}{\bf Exact Tokens} & \multicolumn{3}{c}{\bf Partial Tokens} & \multicolumn{3}{c}{\bf Exact Types} & \multicolumn{3}{c}{\bf Partial Types} \\
\bf Step & \bf Prec & \bf Rec & \bf F1 & \bf Prec & \bf Rec & \bf F1 & \bf Prec & \bf Rec & \bf F1 & \bf Prec & \bf Rec & \bf F1 \\
\midrule
\\
\multicolumn{3}{l}{a) \textsc{Guzman} dataset:} \\
\midrule
Pre-processing & 39.5 & 12.9 & 18.9 & 52.9 & 17.7 & 25.8 & 45.5 & 14.6 & 21.4 & 68.9 & 26.2 & 37.2 \\
Simulation I & 39.6 & 12.6 & 18.6 & 53.3 & 17.4 & 25.5 & 44.5 & 14.0 & 20.7 & 68.1 & 25.4 & 36.1 \\
Simulation II & 39.6 & 11.6 & 17.2 & 48.6 & 14.4 & 21.1 & 43.3 & 13.0 & 19.1 & 66.5 & 23.2 & 33.1 \\
Simulation III-3 & 38.9 & 12.2 & 17.8 & 51.7 & 16.7 & 24.2 & 48.0 & 15.4 & 22.5 & 74.9 & 29.9 & 41.8\\
\midrule
\\
\multicolumn{3}{l}{b) \textsc{Shah-I} dataset:} \\
\midrule
Pre-processing & 75.5 & 48.4 & 57.9 & 80.9 & 51.6 & 61.7 & 60.1 & 27.6 & 37.4 & 80.0 & 42.3 & 55.0 \\
Simulation I & 42.8 & 10.8 & 16.8 & 47.4 & 12.1 & 18.8 & 50.5 & 13.2 & 20.5 & 69.5 & 22.5 & 33.6 \\
Simulation II & 57.6 & 12.1 & 19.7 & 62.2 & 13.2 & 21.5 & 58.6 & 12.9 & 20.9 & 78.8 & 22.1 & 34.1\\
Simulation III-3& 55.2 & 13.5 & 21.4 & 59.5 & 14.7 & 23.3 & 60.2 & 15.2 & 24.1 & 82.0 & 25.8 & 39.0\\
\midrule
\\
\multicolumn{3}{l}{c) \textsc{Shah-II} dataset:} \\
\midrule
Pre-processing  & 33.7 & 11.1 & 16.6 & 38.1 & 12.6 & 18.9 & 46.0 & 14.3 & 21.4 & 64.2 & 21.3 & 31.5 \\
Simulation I & 44.9 & 12.6 & 18.7 & 48.8 & 14.0 & 20.8 & 48.9 & 12.4 & 18.8 & 68.2 & 17.2 & 25.9\\
Simulation II & 38.2 & 14.2 & 19.8 & 41.8 & 15.7 & 21.9 & 40.4 & 13.2 & 19.2 & 56.6 & 17.7 & 25.9\\
Simulation III-3 & 50.2 & 13.9 & 21.0 & 55.0 & 15.6 & 23.5 & 57.8 & 14.2 & 22.3 & 68.0 & 18.5 & 28.4\\
\midrule
\\
\multicolumn{3}{l}{d) \textsc{Sanger} dataset:} \\
\midrule
Pre-processing  & 70.3 & 49.9 & 57.9 & 76.4 & 54.4 & 63.1 & 63.5 & 38.9 & 47.7 & 74.7 & 49.6 & 59.2 \\
Simulation I & 58.9 & 32.8 & 41.6 & 65.3 & 36.5 & 46.3 & 60.8 & 32.1 & 41.4 & 71.8 & 40.7 & 51.3 \\
Simulation II& 52.6 & 31.5 & 39.1 & 59.0 & 35.4 & 43.9 & 54.8 & 32.0 & 40.0 & 66.5 & 40.9 & 50.3 \\
Simulation III-3& 54.5 & 30.4 & 38.6 & 60.1 & 33.7 & 42.7 & 57.2 & 30.6 & 39.5 & 68.5 & 39.0 & 49.3\\
\bottomrule
\end{tabular}
\captionsetup{justification=justified}
\caption{Model performance on all datasets after all data processing steps. 
}
\label{tab:rq1_res}
\end{table}

The second row in each section of Table~\ref{tab:rq1_res} (Simulation I) shows the performance after filtering out app features referring to the app itself. This corresponds to Step 1 of our data processing steps.
It is a simulation of a change in the annotation guidelines, i.e., the explicit mentioning that references to the app itself should not be annotated. The effect on the datasets of retrospectively applying this rule can be seen in Figure~\ref{fig:figure1}. In Step 1, all datasets have a high type-token ratio in the range [0.68, 0.78]. All English datasets have a low share of single-word features in the range [35\%, 38\%], while the one German dataset (\textsc{Sanger}) still has a relatively large portion of single-word features (77\%).
As expected, all results on datasets with previously high performance (\textsc{Shah-I} and \textsc{Sanger}) drop considerably, especially the recall on the \textsc{Shah-I} dataset. 

The third row in each section of Table~\ref{tab:rq1_res} (Simulation II) shows the performance of our models after removing app features that do not contain a noun. This corresponds to Step 2 of our data processing steps.
This step was motivated by the assumption that app features not containing a noun (such as \emph{running} or \emph{runs}) are too unspecific to be useful for the developers. Since the \textsc{Sanger} annotation guidelines instruct to annotate implicit features represented by a single verb, we expected a significant drop of the number of app features for the \textsc{Sanger} and \textsc{Shah} datasets and also an over-proportional reduction of the number of single-word app features. Surprisingly, it turned out that the number of app features dropped equally strongly for the \textsc{Guzman} dataset, and for all datasets the portion of single-word app features also significantly decreased but not much as compared to Step 1 (Simulation I), while the type-token ratio slightly increased. The German dataset \textsc{Sanger} still has a high portion of single-word app features (now 74\%). The average number of app features per review narrows down after this step to the range [0.39,1.42] and is decreasing for all the datasets.

Overall, compared to the performance obtained after Step 1 (Simulation I), the recall remains roughly the same for all datasets. In terms of precision, we expected it to improve. If the annotated feature set contains short and vague verbal aspects that also would be used as non-aspect terms in the text (e.g. \emph{using} or \emph{updating}), it might be very difficult for the model to detect certain instances of these words as features. We expected that removing such features from the annotated set would thus improve precision. The results show, however, that the precision increases only on the \textsc{Shah-I} dataset while on all other datasets it dropped. This can happen if the short and vague-meaning verbal features share the same characteristics with the self-references---the distinct number of such features is small but their frequency is high, in which case it is relatively simple for the model to spot them and removing these features from the annotations causes the precision to drop.

\begin{figure}[]
	\centering
	\includegraphics[width=0.90\textwidth]{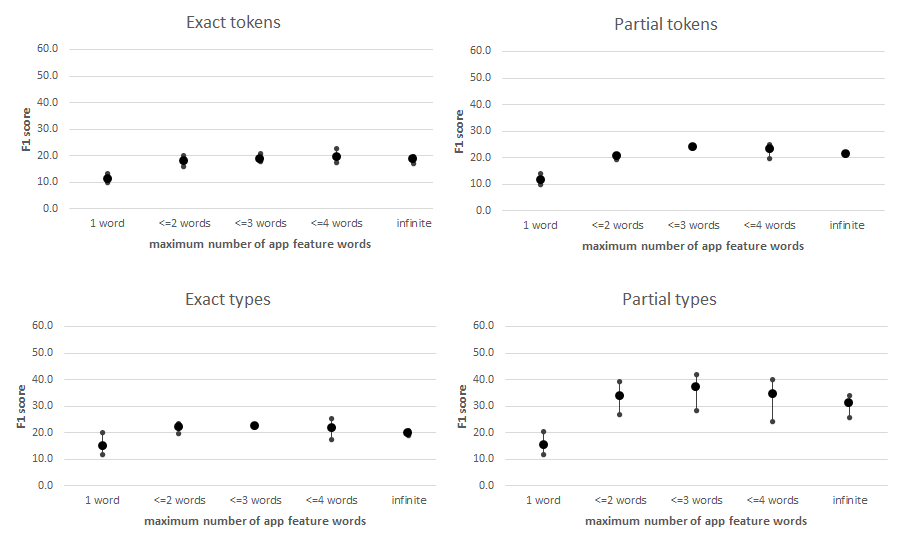}
    \captionsetup{justification=justified}
	\caption{Average f1-score for different evaluation types (exact tokens, partial tokens, exact types and partial types) when applying different cut-offs to the number of words in app features. The results are averaged over three English datasets, showing also minimum and maximum values.}
	\label{fig:cut_off_words_plot}
\end{figure}

The performance results shown for Steps 0 to 2 were achieved with annotations of app features (aspects) consisting of any number of words. 
Since long aspects potentially have a negative effect on model performance, we investigated whether imposing a maximum length could achieve better performance. 
Figure~\ref{fig:cut_off_words_plot} summarizes the outcomes of our experiments. For the \textsc{sanger} dataset, the performance was uniform across all choices of cutoffs. Therefore, we only show the average performances of the three English datasets. Each of the four plots shown in Figure~\ref{fig:cut_off_words_plot} shows the minimum, maximum, and average F1-score for app features containing a number of words not greater than 1, 2, 3, 4 and with infinite length (from left to right). The plot in the upper left corner corresponds to exact token-based evaluation, followed by the plot for partial token-based evaluation (upper right corner), exact type-based evaluation (lower left corner), and partial type-based evaluation (lower right corner).
In three plots out of four, the best average performance is achieved when the app features consisting of more than three words are removed---only in the exact tokens setting restricting the length of app features to four words is slightly better. 
The performance of our models after limiting the number of words in app features to a maximum of three words is shown in the last row of each section in Table~\ref{tab:rq1_res} (Simulation III-3). 
This corresponds to Step 3 of our data processing steps. The effect on performance is uniformly positive for all datasets. The precision for partial type-based evaluation is in the range [68\%,82\%].

Thus, to answer \emph{RQ1}, we can state that by simulating the application of modified annotation guidelines we achieve a feature prediction precision which is comparable to that received after the application of the original \textsc{Guzman} and \textsc{Sanger} guidelines. 
The advantage of the models created based on annotated datasets achieved by simulating the modified guidelines is that the predicted app features are more useful for developers since they are crisper (only one to three words of length) and correlate better with actual app features than with pseudo-features such as references to the app itself. 
The new rules included in the modified (improved) guidelines can be summarized as follows:
\begin{itemize}
\item Only annotate app features consisting of consecutive words;
\item Do not annotate references to the app itself;
\item Only annotate app features containing a noun;
\item Restrict the length of the annotated app features to maximum three words.
\end{itemize}

\subsection{RQ2: How sensitive is automatic feature extraction from app reviews to the size and scope
of the annotated datasets used?}

\begin{table}[]
\centering
\begin{tabular}{l@{\hspace{0.5\tabcolsep}}lrrrrrrrrrrrr}
\toprule
\multicolumn{2}{l}{\bf Procedure} & \multicolumn{3}{c}{\bf Exact Tokens} & \multicolumn{3}{c}{\bf Partial Tokens} & \multicolumn{3}{c}{\bf Exact Types} & \multicolumn{3}{c}{\bf Partial Types} \\
\multicolumn{2}{l}{\bf and size}  & \bf Prec & \bf Rec & \bf F1 & \bf Prec & \bf Rec & \bf F1 & \bf Prec & \bf Rec & \bf F1 & \bf Prec & \bf Rec & \bf F1 \\
\midrule
\\
\multicolumn{3}{l}{a) \textsc{Guzman} dataset:} \\
\midrule
\textsc{appCat} & S & 36.2 & 14.1 & 20.0 & 45.8 & 17.8 & 25.3 & 43.9 & 16.2 & 23.5 & 61.8 & 24.1 & 34.3\\
\textsc{CCV} & M & 38.9 & 12.2 & 17.8 & 51.7 & 16.7 & 24.2 & 48.0 & 15.4 & 22.5 & 74.9 & 29.9 & 41.8\\
\textsc{SCV} & M & 40.6 & 18.3 & 25.2 &52.9 & 23.5 & 32.6 &48.2 & 21.1 & 29.3 &71.1 & 33.9 & 45.9\\
\textsc{CCV-Ext} & L & 24.0 & 17.7 & 19.8 & 34.5 & 25.2 & 28.4 & 31.1 & 22.4 & 25.6 & 55.4 & 43.4 & 48.2\\
\textsc{SCV-Ext} & L & 29.0 & 18.9 & 22.8 &40.9 & 26.6 & 32.2 &36.3 & 23.0 & 28.2 &59.1 & 38.8 & 46.9\\
\midrule
\\
\multicolumn{3}{l}{b) \textsc{Shah-I} dataset:} \\
\midrule
\textsc{appCat} & S & 43.4 & 13.0 & 19.8 & 47.0 & 14.6 & 22.0 & 47.4 & 14.4 & 21.9 & 52.8 & 19.6 & 28.2\\
\textsc{CCV} & M & 55.2 & 13.5 & 21.4 & 59.5 & 14.7 & 23.3 & 60.2 & 15.2 & 24.1 & 82.0 & 25.8 & 39.0\\
\textsc{SCV} & M & 66.2 & 30.4 & 41.7 &70.5 & 32.4 & 44.4 &68.5 & 30.6 & 42.3 &81.8 & 38.8 & 52.7\\
\textsc{CCV-Ext} & L & 17.9 & 21.8 & 19.2 & 21.8 & 26.4 & 23.2 & 24.5 & 30.7 & 26.9 & 40.6 & 52.1 & 45.3\\
\textsc{SCV-Ext} & L & 31.4 & 28.3 & 29.8 &34.9 & 31.6 & 33.2 &34.4 & 31.6 & 32.9 &45.1 & 43.2 & 44.2 \\
\midrule
\\
\multicolumn{3}{l}{c) \textsc{Shah-II} dataset:} \\
\midrule
\textsc{appCat} & S & 18.2 & 5.7 & 8.6 & 21.6 & 6.9 & 10.2 & 24.3 & 7.8 & 11.7 & 29.6 & 9.3 & 14.0\\
\textsc{CCV} & M & 50.2 & 13.9 & 21.0 & 55.0 & 15.6 & 23.5 & 57.8 & 14.2 & 22.3 & 68.0 & 18.5 & 28.4\\
\textsc{SCV} & M & 39.7 & 13.6 & 20.2 &46.8 & 16.1 & 24.0 &50.9 & 15.8 & 24.1 &65.6 & 21.1 & 31.9\\
\textsc{CCV-Ext} & L & 16.8 & 21.4 & 18.6 & 20.2 & 25.8 & 22.5 & 21.6 & 24.5 & 22.8 & 34.6 & 40.2 & 37.0\\
\textsc{SCV-Ext} & L & 28.3 & 25.3 & 26.7 &32.0 & 28.3 & 30.0 &29.2 & 24.8 & 26.8 &37.6 & 31.8 & 34.4\\
\midrule
\\
\multicolumn{3}{l}{d) \textsc{Sanger} dataset:} \\
\midrule
\textsc{appCat} & S & 56.3 & 21.0 & 30.3 & 63.4 & 23.6 & 33.9 & 59.4 & 22.5 & 32.3 & 66.1 & 26.0 & 36.9\\
\textsc{CCV} & M & 54.5 & 30.4 & 38.6 & 60.1 & 33.7 & 42.7 & 57.2 & 30.6 & 39.5 & 68.5 & 39.0 & 49.3\\
\textsc{SCV} & M & 56.0 & 34.8 & 42.9 &62.9 & 39.2 & 48.3 &55.2 & 32.9 & 41.2 &65.4 & 39.9 & 49.6\\
\bottomrule
\end{tabular}
\captionsetup{justification=justified}
\caption{ Results of various training procedures that differ in size and scope on all datasets. The second column shows the size of the training set in procedure according to a three-value scale: small (S), medium (M) and large (L). \textsc{appCat} stands for the cross-validated training procedure trained on one app category data only; \textsc{CCV} means Cross-Category Validation; \textsc{SCV} is Stratified Cross Validation; \textsc{CCV-Ext} is Cross-Category Validation with External data; and \textsc{SCV-Ext} is Stratified Cross Validation with External data.}
\label{tab:rq2_res}
\end{table}

In order to investigate how sensitive the performance trained on the datasets annotated with simulated annotation guidelines (cf. rows Simulation III-3 in Table~\ref{tab:rq1_res})
is regarding to variations in size and scope of available annotated datasets is, we compared five training procedures: 
\begin{enumerate*}[label=\arabic*)]
\item Cross-category validation (\textsc{CCV}),
\item 10-fold cross-validation over single app category (\textsc{appCat},
\item Stratified 10-fold cross-validation (\textsc{SCV}),
\item Cross-category validation with external \textsc{Laptop} and \textsc{Restaurant} datasets (\textsc{CCV-Ext})
\item Stratified 10-fold cross-validation over merged app categories with external \textsc{Laptop} and \textsc{Restaurant} datasets (\textsc{SCV-Ext}).
\end{enumerate*}
These training procedures were described in subsection~\ref{sec:training_procedures}.

The five training procedures can be classified with regards to size and scope as follows:
\begin{itemize}
\item Small size and scope: \textsc{appCat} -- scope: reviews from one app category; size: number of app features (tokens) in the range [17, 605];
\item Medium size and scope: \textsc{CCV} and \textsc{SCV} -- scope: reviews from all app categories; size: number of app features (tokens) in the range [486, 1435];
\item Large size and scope: \textsc{CCV-Ext} and \textsc{SCV-Ext} -- scope: reviews from all app categories plus \textsc{Laptop} and \textsc{Restaurant}; size: number of app features (token) in the range [8325, 9274].
\end{itemize}

\begin{figure}[]
	\centering
	\includegraphics[width=1.0\textwidth]{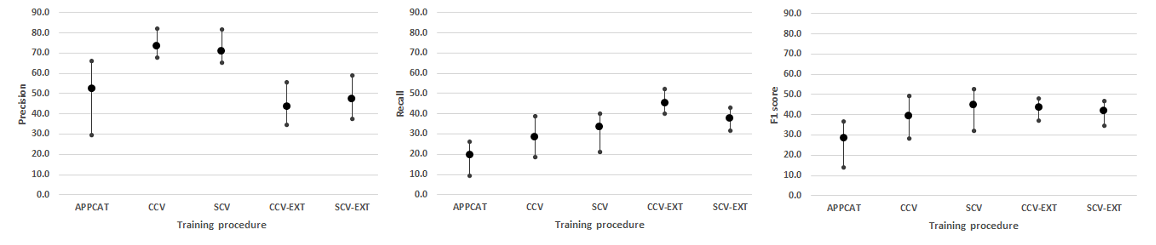}
    \captionsetup{justification=justified}
	\caption{Average precision, recall and f1-score for partial types evaluation when applying different training procedures.}
	\label{fig:figure3}
\end{figure}

The performances of the models resulting from the different training procedures are shown  in Table~\ref{tab:rq2_res}.
Figure~\ref{fig:figure3} shows in a more compact representation the minimum, maximum, and average f1-score for the five training procedures for the various datasets using partial type-based evaluation procedure. The other evaluation metrics follow the same pattern.

As expected, all models using the \textsc{appCat} training procedure have the lowest performance for all evaluation measures. The most probable explanation for this result is the small size of the training sets as each model is trained on the annotated reviews of one app category only. 

The best performing models with regard to precision are those resulting from the training procedures \textsc{CCV} and \textsc{SCV}. The training procedure \textsc{CCV} is significantly better in terms of precision than \textsc{SCV}. In terms of recall, the procedure \textsc{SCV} is consistently better than \textsc{CCV} on all datasets, leading also to consistently higher f1-scores. However, this improvement is bought with the necessity to have annotated app reviews available in all app categories, while \textsc{CCV} procedure applies the trained model on reviews from app categories that were not seen during training time.


Since the \textsc{Laptop} and \textsc{restaurant} datasets are in English, they cannot be combined with the \textsc{Sanger} dataset and we only have performance evaluations including those external datasets for \textsc{Guzman} and both \textsc{Shah} datasets. The evaluation results show that widening the scope and using non-app reviews as additional training data improves the recall but reduces precision (with an improvement of the f1-score) when comparing \textsc{CCV-Ext} to \textsc{CCV}. Comparing \textsc{SCV-Ext} to \textsc{SCV} shows that adding external datasets improves recall but as the drop in precision is in most cases large, the f1-score of \textsc{SCV-Ext} increases less than in the case of \textsc{CCV} (\textsc{Guzman} and \textsc{Shah-II} datasets) or even drops (\textsc{Shah-I} dataset).

Thus, to answer \emph{RQ2}, we can state that to extract features from a particular app category, the training procedure \textsc{CCV}, which uses annotated reviews from other app categories, achieves the highest precision. On the other hand, \textsc{SCV}, the training procedure that uses reviews from the same app category, yields better recall at the cost of a drop in precision. In addition, we can say that using annotated app reviews from other app categories to increase the training set does improve the overall performance (\textsc{appCat} vs \textsc{CCV/SCV}). While complementing annotated app review datasets with external datasets from different domains helps to increase recall, it also brings along a large drop in precision. Since software developers usually would be interested in more precise app feature predictions (at the cost of lower recall), dataset extension by adding n`on-app features might not be the recommended direction to follow.
 
%% file: discussion.tex
\section{Discussion}
\label{sec:discussion}

In this section, we first explain the value of our proposed new annotation guidelines (AGs). Then we discuss limitations of our study.

\subsection{Usefulness of the New Annotation Guidelines}
\label{sec:ag_usefulness}

The main goal of our study was to investigate the impact of annotation guidelines and annotated data on extracting app features from app reviews and to improve existing AGs such that (1) the performance of the app feature extraction task gets better in terms of f1 score and (2) the set of extracted app features is more useful to software developers. 

Section~\ref{sec:results} (Results) presented the step-by-step impact to the performance of the app feature extraction when simulating the effects of changing the used AGs.
It turned out that with our proposed new AGs, a small performance improvement over the baseline situation could be achieved. However, this is not the only advantage of our new AGs. In the following we argue that not only the performance of the app feature extraction task can be improved but that the set of annotated and extracted app features itself is more relevant for software developers when using our new AGs.


We will illustrate what we mean by ``more relevant for software developers'' in two steps with the help of examples. In the first step, we demonstrate that the simulated application of our new AGs actually produces an annotated dataset that contains a larger share of annotated app features that are useful to software developers. In the second step, we demonstrate that this positive effect of our new annotation guidelines also propagates to the set of extracted app features.

Table~\ref{tab:features_before_after_AGs} shows samples of app features in the original labeled datasets and in the annotated datasets after the simulated application of our new AGs. We randomly picked one app category from each of the English datasets, i.e., in categories `Photography' (\textsc{Guzman}), `Social' (\textsc{Shah-I}), and `Game' (\textsc{Shah-II}). We manually classified each app feature as either `useful' or `not useful' and then compared how the numbers of useful and not useful app features change when simulating the application of our new AGs. In Table~\ref{tab:features_before_after_AGs}, not useful app features are shown in bold text.

We consider an app feature to be useful when it seems to be related to actual functionality of an app. For example \emph{capture full resolution}, \emph{decorating pictures} and \emph{online scrapbooking} seem to be clearly referencing to some functionality in the app of categories `Photography' and `Social'. Aspects are not useful when they are too generic to be connected with a specific functionality (e.g., \emph{share} or \emph{version 1.5.1}) or when they are referring to a non-functional aspect of an app (e.g., \emph{easy to use}). 

The app features in the upper part of Table~\ref{tab:features_before_after_AGs} correspond to a random sample of those app features that remained in the set of app features after simulated application of the new AGs. The numbers behind each app feature correspond to the token count before and after the simulated application of the new AGs. Note that in some cases, the token number changed. The app features in the lower part of Table~\ref{tab:features_before_after_AGs} correspond to those app features that were completely removed from the set of app features after the simulated application of the new AGs. Again, Table~\ref{tab:features_before_after_AGs} only shows a random sample of the removed app features.

\begin{table}[!ht]

\small
\caption{A random sample of app features from three app categories: photography (\textsc{guzman}), social (\textsc{shah-I}), and game (\textsc{shah-II}). The upper part of the table shows app features kept or partially removed during the simulation of our AGs; while the lower part shows the app features removed by the AGs. The app features shown in bold are not useful from developer's prospective.}
\label{tab:features_before_after_AGs}
\resizebox{\textwidth}{!}{%
\begin{tabular}{l|l|l}
\midrule
\textbf{Photography (\textsc{Guzman})} & \textbf{Social (\textsc{Shah-I})}  & \textbf{Game (\textsc{Shah-II})}  \\
\midrule
editing [13,5] & pinning [7,4] & new levels [12,12] \\
 \textbf{edit [9,4]} & \textbf{ideas [5,3]} & \textbf{most recent update [1,1]} \\
 \textbf{easy to use [6,3]} & \textbf{works [4,1]} & \textbf{new version [1,1]} \\
 \textbf{free [3,1]} & \textbf{update [2,2]} & sound [1,1]\\
 filters [3,3] & block pins[1,1] & new pink bird [1,1]\\
 photoshop [2,2] & login [1,1] & \textbf{version 1.5.1 [1,1]} \\
\textbf{effects [1,1]} & pin videos [1,1] & new gameplay [1,1] \\
capture full resolution [1,1] & search results [1,1] & indigo bird [1,1] \\
\textbf{free pack [1,1]} & share interests [1,1] & new powerups [1,1] \\
enhancing photos [1,1] & move pin button [1,1] & tapping and aiming [1,1] \\
more frames [1,1] & online scrapbooking [1,1] & red's might feathers [1,1] \\
\textbf{share [1,1]} & \textbf{uploading [1,1]} & trajectory line [1,1] \\
customize a photo [1,1] & find pins [1,1] & pop ups [1,1]\\
\textbf{nexus 5} [1,1] & \textbf{explore [1,1]} & physics engine [1,1] \\
textfonts [1,1] & organize your interests [1,1] & \textbf{newest update} [1,1] \\
decorating pictures [1,1] & search function [1,1] & new format [1,1] \\
edit pics [1,1] & send pins [1,1] & turn off cartoons [1,1] \\
customised jpeg file [1,1] & image searcher [1,1] & mighty eagle [1,1] \\
add more frame [1,1] & change cover photo [1,1] & tower defense [1,1] \\
red eye optn [1,1] & random pin button [1,1] & birds reload [1,1] \\
\midrule
\textbf{update [1,0]} & \textbf{app [140,0]} & \textbf{game [26,0]} \\
\textbf{explore [1,0]} & \textbf{pinterest [42,0]} & \textbf{angry birds [7,0]} \\
\textbf{save [1,0]}  & \textbf{apps [3,0]} & \textbf{app [4,0]} \\
\textbf{upload [1,0]} &  pin [2,0]  & \textbf{play [3,0]} \\
\textbf{download [1,0]} & \textbf{runs [2,0]} & \textbf{games [1,0]} \\
\textbf{user friendly [1,0]} & \textbf{application [1,0]}  & \textbf{delete[1,0]} \\
\textbf{little slow [1,0]} & \textbf{navigate} [1,0]& \textbf{most of the levels [1,0]} \\
\textbf{edited [1,0]} & \textbf{quotes [1,0]} & \textbf{runs [1,0]} \\
\textbf{navigate [1,0]} & \textbf{loads [1,0]} & \textbf{open [1,0]}\\
add more clip arts [1,0] &  view someone elses board [1,0] & icould support [1,0] \\
add other output formats [1,0] &  change the privacy settings on boards [1,0] & make your own levels [1,0]  \\
edit pictures in a high quality [1,0] & update on the news feed [1,0] & replay an awesome shot [1,0] \\
\textbf{takes forever to apply effects [1,0]} & sorting functionality in board section [1,0] & more levels with pink bird [1,0] \\
selection tool to edit and work & making sub folder or subcategories & super hero angry birds [1,0] \\
with image parts [1,0] & within boards [1,0] &  \\
\midrule
\end{tabular}}
\end{table}

\begin{table}[!ht]
\centering
\small
\caption{The number of app features (token and type) manually classified as either `useful' or `not useful' in app categories Photography (\textsc{guzman}), Social (\textsc{shah-I}) and Game (\textsc{shah-II}), before and after the simulation of AGs.}
\label{tab:useful_vs_useless_appFeatures} 
\resizebox{\textwidth}{!}{%
\begin{tabular}{llcccccc}
\toprule
\textbf{App} & \textbf{Before} & \textbf{Useful} & \textbf{Not useful} & \textbf{Useful} & \textbf{Not useful} & \textbf{Total} & \textbf{Total}\\
\textbf{category} & \textbf{or}  & \textbf{app features} & \textbf{app features} & \textbf{app features} & \textbf{app features} & \textbf{app features} & \textbf{app features}\\
 & \textbf{after simulation} & \textbf{(token)} & \textbf{(token)} &  \textbf{(type)} & \textbf{(type)} & \textbf{(token)} & \textbf{(type)} \\
\toprule
\multirow{2}{*}{Photography} & Before simulation & 79 & 95 & 50 & 77 & 174 & 127 \\
 &  After simulation & 55 & 45 & 37 & 41 & 100 & 78\\
 \midrule
 \multirow{2}{*}{Social} & Before simulation & 69 & 217 & 47 & 19 & 286 & 66\\
 &  After simulation & 52 & 15 & 36 & 8 & 67 & 44\\
 \midrule
  \multirow{2}{*}{Game} & Before simulation & 81 & 59 & 57 & 22& 140 & 79\\
 & After simulation & 73 & 12 & 49 & 13 & 85 & 62\\
 \bottomrule
\end{tabular}}
\end{table}

The ideal impact of the simulated application of our new AGs corresponds to removing all useless app features and keeping only the useful app features. We calculated the impact of our AGs based on the numbers of manually classified `useful' and `not useful' app features in three app categories before and after the simulation of new AGs (see Table~\ref{tab:useful_vs_useless_appFeatures}). The actual numbers (based on token and type count) for each of the three apps are as follows:
\begin{itemize}
\item Category `Photography' (\textsc{Guzman}): the token and type percentage of useful app features kept equals 70\% and 53\%, respectively; the token and type percentage of useless app features removed equals 53\% and 47\%, respectively; the ratio between useful and useless app features improved from 79/95=0.83 (token-based) and 50/77=0.65 (type-based) before the application of our new AGs to 55/45=1.22 (token-based) and 37/41=0.90 (type-based) afterwards; 
\item Category `Social' (\textsc{Shah-I}): the token and type percentage of useful app features kept equals 75\% and 77\%, respectively; the percentage of useless app features removed equals 93\% and 58\%, respectively; the ratio between useful and useless app features improved from 69/217=0.32 (token-based) and 47/19=2.47 (type-based) before the application of our new AGs to 52/15=3.47 (token-based) and 36/8=4.50 (type-based) afterwards; 
\item Category `Game' (\textsc{Shah-II}): the token and type percentage of useful app features kept equals 90\% and 86\%, respectively; the percentage of useless app features removed equals 80\% and 41\%, respectively; the ratio between useful and useless app features improved from 81/59=1.37 (token-based) and 57/22=2.59 (type-based) before the application of our new AGs to 73/12=6.08 (token-based) and 49/13=3.77 (type-based) afterwards. 
\end{itemize}

The data shows for each of the three sample cases that the ratio between the number of useful and useless app features is increasing for both token and type-based analyses when applying our new AGs. This is the effect that we expected to see.

We computed the percentages based on both token and type counts of app features because there can be cases like, for example, the app feature \emph{editing}. The app feature \emph{editing} occurred 13 times in the app of category `Photography' before the simulated application of our new AGs and five times afterwards. We assume that eight occurrences of \emph{editing} were removed due to the guideline `Only annotate app features containing a noun', i.e., because after simulating the application of our new AGs \emph{editing} was predicted to be an app feature when it was used as a noun. Note that the word \emph{editing} when used as a verb is not helpful for software developers because it does not provide information about the purpose or object of editing and thus it is difficult to decide whether the mentioning of \emph{editing} is related to the edit functionality as such or just a special situation in which something was edited. On the other hand, if \emph{editing} is mentioned in the grammatical form of noun, it is more probable that whatever is said in the sentence containing the word \emph{editing} is referring to the edit functionality in general. A similar case is \emph{pinning} mentioned in the reviews of the app in category `Social'. Here three of the seven original app feature predictions disappeared after simulating the application of our new AGs. 

After we convinced ourselves that the simulated application of new AGs actually results in more useful app feature annotations, we checked whether this effect also propagates to set of extracted app features.

Table~\ref{tab:useful_vs_useless_appFeatures_predicted} shows the impact on the number of useful and useless app features in model's extracted app features, when training CRF models with the original annotated datasets and when training CRF models using the annotated datasets after the simulated application of our new AGs. We picked the same app categories as before from each of the English datasets, i.e., from categories 'Photography' (\textsc{guzman}), 'Social' (\textsc{Shah-I}), and 'Game' (\textsc{Shah-II}). We manually classified each app feature as either 'useful' or 'not useful' and then compare how the numbers of useful and not useful app features change when simulating the application of our new AGs. The actual numbers (based on token and type count) for each of the three apps are as follows:
\begin{itemize}
\item Category `Photography' (\textsc{Guzman}): the ratio between useful and useless app features improved from 30/31=0.97 (token-based) and 21/25=0.84 (type-based) before the application of our new AGs to 24/10=2.4 (token-based) and 17/10=1.7 (type-based) afterwards; 
\item Category `Social' (\textsc{Shah-I}): the ratio between useful and useless app features improved from 26/167=0.16 (token-based) and 17/18=0.94 (type-based) before the application of our new AGs to 15/7=2.14 (token-based) and 8/6=1.33 (type-based) afterwards; 
\item Category `Game' (\textsc{Shah-II}): the ratio between useful and useless app features improved from 22/27=0.81 (token-based) and 11/21=0.52 (type-based) before the application of our new AGs to 24/18=1.33 (token-based) and 13/14=0.93 (type-based) afterwards. 
\end{itemize}

\begin{table}[!ht]
\centering
\small
\caption{Model's extracted app features (token and type) are manually classified as either `useful' or `not useful' in app categories Photography (\textsc{guzman}), Social (\textsc{shah-I}) and Game (\textsc{shah-II}), before and after the simulation of AGs.}
\label{tab:useful_vs_useless_appFeatures_predicted} 
\resizebox{\textwidth}{!}{%
\begin{tabular}{llcccccc}
\toprule
\textbf{App} & \textbf{Before} & \textbf{Useful} & \textbf{Not useful} & \textbf{Useful} & \textbf{Not useful} & \textbf{Total} & \textbf{Total}\\
\textbf{category} & \textbf{or}  & \textbf{app features} & \textbf{app features} & \textbf{app features} & \textbf{app features} & \textbf{app features} & \textbf{app features}\\
 & \textbf{after simulation} & \textbf{(token)} & \textbf{(token)} &  \textbf{(type)} & \textbf{(type)} & \textbf{(token)} & \textbf{(type)} \\
\toprule
\multirow{2}{*}{Photography} & Before simulation & 30 & 31 & 21  & 25 & 61 & 46 \\
 &  After simulation & 24 & 10 & 17 & 10 & 27 & 34\\
 \midrule
 \multirow{2}{*}{Social} & Before simulation & 26 & 167 & 17 & 18 & 193 & 35\\
 &  After simulation & 15 & 7 &  8 & 6 & 22 & 14 \\
 \midrule
  \multirow{2}{*}{Game} & Before simulation & 22 & 27 & 11 & 21 & 49 & 32\\
 & After simulation & 24 & 18 & 13 & 14 & 42 & 27\\
 \bottomrule
\end{tabular}}
\end{table}

\begin{figure}[!ht]
\centering
\includegraphics[width=0.90\linewidth]{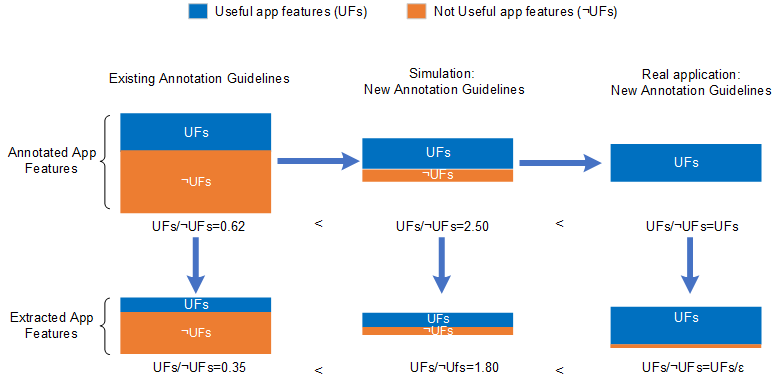}
\caption{Ratios between useful and useless app features (annotated and extracted) for the three analyzed app categories (token-based).}
\label{fig:schematic_diagram}
\end{figure}

Figure~\ref{fig:schematic_diagram} summarizes our results and expectations with regards to the effectiveness of our new AGs (based on token-wise analysis of the three selected app categories). Each of the six rectangles corresponds to the total set of annotated (upper row) and extracted (lower row) app features. The blue portion in each rectangle corresponds to the share of useful app features (UFs) while the orange portion corresponds to the share of useless app features ($\neg$UFs). The calculated ratios between UFs and $\neg$UFs clearly show an improvement for the simulated application of our new AGs not only in the annotated datasets but also in the set of extracted features. This strengthens our expectation that a real application of the new AGs, which presumably yields exclusively useful features in the annotated dataset (thus an exclusively blue rectangle on the right in the upper row of Figure~\ref{fig:schematic_diagram}) would result in an even further improved ratio between UFs and $\neg$UFs in the set of extracted app features when comparing to the baseline and the simulated application of our new AGs (as indicated by the small portion of orange in the rectangle on the right in the lower row of Figure~\ref{fig:schematic_diagram}).

\subsection{Limitations of the Study}
\label{sec:study_limitations}

In some cases it is not fully clear why an app feature was removed or kept. These cases could be due to inaccuracies of the POS tagger used in Step 2 of the simulated application of our new AGs. For example, it is unclear why only two out of three occasions of \emph{free} were removed in category `Photography' as it is hard to imagine a context in which \emph{free} could be considered to be a noun in a review text. Overall, our new AGs removed most of the useless app features in app categories `Social' (\textsc{Shah-I}) and `Game' (\textsc{Shah-II}). When removing the not useful app features, the lower performance (53\%) on app category `Photography' (\textsc{Guzman}) is due to a large number of annotations referring to mobile devices, app versions, app updates and non-functional app features.

Note that we only simulated the application of our new AGs on the labeled datasets. We expect that the application of the new AGs by actual people could have resulted in more useful annotations of app features in the first place. The application of our new AGs automatically removes app features that are longer than 3-words. However, in the direct application of our new AGs, a longer app feature might simply have been annotated with fewer words rather than completely been removed as we did in Simulation Step 3. For instance, a 5-word app feature \emph{sorting functionality in board section} annotated in app category `Social' of the \textsc{Shah-I} dataset could be labeled as an admissible 2-word app feature \emph{sorting functionality}.

Our study is restricted to the use of the CRF model which limits app features to be annotated as consecutive words. Therefore, when limiting annotations to a maximum 3-word app features, it might be impossible to annotate app features consisting of consecutive words;
in such cases CRF (or any other sequences tagging model) cannot be applied or we would have to drop those app features (or soften the rule of having maximum 3-words app features). For instance, a 5-word app feature \emph{edit pictures in a high quality} can be reduced to the following two meaningful representations of 3-word app features: \emph{edit high quality} or \emph{edit picture quality}. However, both 3-word app feature representations are not consecutive.
 
We only found two published annotation guidelines associated with publicly available annotated app review datasets. The problem we encounter is that either annotated datasets were not published or when they had been published it is unclear what annotation rules/guidelines were applied. In other domains, e.g., \textsc{Laptop} and \textsc{Restaurant},  the standard guidelines and benchmark datasets are contributed by the research community \textsc{SemEval} to perform the task of product feature extraction and its evaluation. Similar to the \textsc{SevEmal} research community, the app review mining research community could contribute standardized guidelines and benchmark datasets to help researchers in the development of systems performing fine-grained sentiment analysis at app feature-level.

We created two new labeled datasets \textsc{Shah-I} and \textsc{Shah-II} in English, using the English translated version of the German \textsc{Sanger} annotation guidelines. The translation from German to English of \textsc{Sanger} guidelines is performed using Google translation service. In order to make sure that the translated guidelines have sufficient quality to be used for the annotation of reviews in English language, one of the author of this paper, who is a native German speaker and have a full command of the English language, read the English translated version of the annotation guidelines and found it adequately accurate for the annotation task.

The validity of our results depends to some degree on the reliability of the annotations of the  \textsc{Shah-I} and \textsc{Shah-II} datasets. In addition, the assessment of the usefulness of the results produced when using our simulated AGs depends on the reliability of the subjective classification of annotated and extracted app features into `useful' and `not useful' for software developers. Since each of these tasks was performed by one person, reliability might be limited. However, since we not only used our own annotations (i.e., datasets  \textsc{Shah-I} and \textsc{Shah-II}) but applied our analyses also to datasets published in the literature and the trend of our results was similar for all our used datasets, we believe that the existing limitations of reliability for the mentioned tasks is not a major threat to validity of our results.


%% file: conclusion.tex
\section{Conclusion}
\label{sec:conclusion}


Previously, several techniques have been used for automatic app feature extraction from app reviews, including
\begin{enumerate*}[label=(\alph*)]
\item unsupervised topic modeling,
\item rule-based methods,
\item supervised machine learning approaces.
\end{enumerate*}
While unsupervised and rule-based approaches only require annotated data for evaluation purposes, supervised machine learning methods also need it for training the model. In either case, the quality of the annotations can considerably affect the evaluation results. When the annotations contain many long complex app features that are difficult for any system to extract then the evaluation results will be artificially low, especially when these features are infrequent and thus not of much actual interest for the app developers. On the other hand, when the annotations contain many short and frequent app features that are easy to detect but not too informative for the developers then the evaluation results will be artificially high. 
Moreover, previous studies in app feature extraction have used different review datasets, annotation guidelines, feature extraction methods and/or evaluation methods, which makes the results of these different studies uncomparable.


In this paper, we focused on two aspects---annotation guidelines and the size and scope of the training data---by controlling the other design parameters as much as possible. We used four different labeled datasets annotated with two different annotation guidelines. For the app feature extraction technique we adopted the supervised CRF method. 
To our best knowledge, this is the first study that explores the impact of annotation guidelines and labeled datasets for app feature extraction from user reviews. As a result of our study, we propose several changes to the existing AGs to avoid the annotation of useless app features. Secondly, our study examined how the size and scope of the training data affects the performance of the app feature extraction.


The first research question, RQ1, started from the observation that the app feature annotations and thus subsequently also automatic feature extraction results varied considerably on different datasets, even though they used the same AGs to annotate the app features and even when the annotated app reviews themselves were identical (\textsc{Shah-I} vs \textsc{Shah-II}).
We hypothesised that these differences are at least partly due to the annotation guidelines that were used to label these datasets. We proposed several changes to the annotation guidelines and evaluated the effect of their simulated application using the evaluation results of the CRF modeling. The proposed changes in AGs include 
\begin{enumerate*}[label=(\alph*)]
\item instructing to annotate only consecutive words as app features,
\item discouraging the annotations to the references of the app itself,
\item instructing to annotate only noun phrases as app features, i.e. every app feature must contain a noun, and
\item instruct to annotate app features with the length of maximum three words.
\end{enumerate*}
When applying the simulated guidelines, we where able to retain the precision of the app feature extraction. However, as after simulation the annotated features in both training and test sets are more informative and less noisy, the modeling result now better reflects the real app feature extraction performance.

In RQ2, we experimented with several ways of using annotated data, exploring whether the training set should include annotated app reviews from test app categories or if the linguistic patterns indicating app features are general enough so that the model trained on the reviews of one set of app categories can then be applied to successfully extract app features from new app categories. We found that in general, it is not necessary to have annotated training data in the test app categories. However, having annotated app reviews in the training set from the test app category enables to improve the recall at the cost of the drop in precision. Additionally, we explored whether utilising additional training data in the form of annotated product reviews would help to improve the performance of app feature extraction. We found that while adding external training data helps to improve the recall, it causes a substantial drop in precision.

The main limitation of our study is the simulation of the new AGs on the labeled review datasets which resulted in removing a number of app features rather than reformulating them according to the new guidelines.
Regardless of that, we believe that the real application of the new AGs by human annotators would have produced a set of app features that are useful and refer to the functional aspects of an app. 
However, this hypothesis is yet to be confirmed by evaluating the proposed AGs by giving them to real human annnotators for labeling app features in user reviews.

%% file: appendix_A.tex
\section{Detailed results of the experiment 'CCV'}
\subsection{\textsc{Guzman} Dataset}

\begin{table}[H]
\centering
\small
\begin{tabular}{lcccccccccccc}
\toprule
\bf App & \multicolumn{3}{c}{\bf Exact Tokens} & \multicolumn{3}{c}{\bf Partial Tokens} & \multicolumn{3}{c}{\bf Exact Types} & \multicolumn{3}{c}{\bf Partial Types} \\
\bf Category & \bf Prec & \bf Rec & \bf F1 & \bf Prec & \bf Rec & \bf F1 & \bf Prec & \bf Rec & \bf F1 & \bf Prec & \bf Rec & \bf F1 \\
\midrule
\multicolumn{3}{l}{a) Pre-processing:} \\
\midrule
Game & 46.8 & 8.8 & 14.8 &64.5 & 12.1 & 20.4 &54.1 & 8.9 & 15.3 &79.2 & 18.8 & 30.3\\
Productivity & 29.2 & 16.3 & 20.9 &40.2 & 22.5 & 28.9 &40.3 & 20.2 & 26.9 &69.4 & 40.0 & 50.7\\
Travel & 49.6 & 13.5 & 21.2 &61.1 & 16.6 & 26.1 &57.7 & 15.7 & 24.7 &78.7 & 25.8 & 38.8\\
Photography & 36.1 & 13.8 & 19.9 &52.5 & 20.0 & 29.0 &37.8 & 16.2 & 22.7 &59.1 & 24.8 & 34.9\\
Social & 51.0 & 12.6 & 20.2 &58.8 & 14.5 & 23.3 &60.0 & 14.1 & 22.8 &78.6 & 22.1 & 34.6\\
Communication & 24.6 & 12.4 & 16.5 &40.4 & 20.4 & 27.1 &23.1 & 12.5 & 16.2 &48.1 & 26.0 & 33.8\\
\bottomrule
\textbf{Average} & \textbf{39.5} & \textbf{12.9} & \textbf{18.9} & \textbf{52.9} & \textbf{17.7} & \textbf{25.8} & \textbf{45.5} & \textbf{14.6} & \textbf{21.4} & \textbf{68.9} & \textbf{26.2} & \textbf{37.2}\\
\bottomrule

\midrule
\multicolumn{3}{l}{b) Simulation I:} \\
\midrule
Game & 48.3 & 8.5 & 14.4 &67.2 & 11.8 & 20.1 &51.4 & 8.0 & 13.9 &76.6 & 16.1 & 26.6\\
Productivity & 29.2 & 15.8 & 20.5 &40.9 & 22.1 & 28.7 &40.8 & 19.9 & 26.8 &70.2 & 39.5 & 50.6\\
Travel & 50.0 & 13.6 & 21.4 &60.7 & 16.5 & 26.0 &57.1 & 15.4 & 24.3 &78.3 & 25.3 & 38.2\\
Photography & 33.9 & 12.5 & 18.3 &50.8 & 18.8 & 27.4 &34.1 & 14.3 & 20.1 &55.8 & 22.9 & 32.4\\
Social & 51.0 & 12.6 & 20.2 &58.8 & 14.5 & 23.3 &60.0 & 14.1 & 22.8 &78.6 & 22.1 & 34.6\\
Communication & 25.0 & 12.7 & 16.9 &41.1 & 20.9 & 27.7 &23.5 & 12.6 & 16.4 &49.0 & 26.3 & 34.2\\
\bottomrule
\textbf{Average} & \textbf{39.6} & \textbf{12.6} & \textbf{18.6} & \textbf{53.3} & \textbf{17.4} & \textbf{25.5} & \textbf{44.5} & \textbf{14.0} & \textbf{20.7} & \textbf{68.1} & \textbf{25.4} & \textbf{36.1}\\
\bottomrule

\midrule
\multicolumn{3}{l}{c) Simulation II:} \\
\midrule
Game & 53.7 & 9.6 & 16.3 &64.8 & 11.6 & 19.7 &53.1 & 8.2 & 14.2 &79.5 & 15.0 & 25.2\\
Productivity & 26.1 & 19.3 & 22.2 &35.6 & 26.4 & 30.3 &31.7 & 22.3 & 26.2 &59.0 & 42.8 & 49.6\\
Travel & 42.0 & 8.3 & 13.9 &55.1 & 10.9 & 18.2 &44.3 & 10.8 & 17.3 &72.9 & 24.7 & 36.9\\
Photography & 31.2 & 8.8 & 13.7 &37.5 & 10.5 & 16.4 &37.0 & 11.6 & 17.7 &51.9 & 16.3 & 24.8\\
Social & 43.2 & 10.7 & 17.1 &56.8 & 14.0 & 22.5 &51.9 & 11.6 & 18.9 &74.2 & 19.0 & 30.3\\
Communication & 41.7 & 13.0 & 19.8 &41.7 & 13.0 & 19.8 &41.7 & 13.3 & 20.2 &61.5 & 21.3 & 31.7\\
\bottomrule
\textbf{Average} & \textbf{39.6} & \textbf{11.6} & \textbf{17.2} & \textbf{48.6} & \textbf{14.4} & \textbf{21.1} & \textbf{43.3} & \textbf{13.0} & \textbf{19.1} & \textbf{66.5} & \textbf{23.2} & \textbf{33.1}\\
\bottomrule

\midrule
\multicolumn{3}{l}{d) Simulation III-3:} \\
\midrule
Game & 46.0 & 11.4 & 18.3 &66.7 & 16.5 & 26.5 &62.9 & 14.3 & 23.3 &92.6 & 32.5 & 48.1\\
Productivity & 26.5 & 19.0 & 22.1 &38.8 & 27.8 & 32.4 &37.8 & 24.5 & 29.7 &78.0 & 55.1 & 64.6\\
Travel & 53.5 & 8.0 & 13.9 &60.5 & 9.0 & 15.7 &61.1 & 11.5 & 19.4 &88.5 & 28.3 & 42.9\\
Photography & 26.5 & 9.3 & 13.7 &41.2 & 14.4 & 21.4 &38.5 & 14.5 & 21.1 &68.0 & 24.6 & 36.2\\
Social & 43.6 & 14.3 & 21.5 &59.0 & 19.3 & 29.1 &50.0 & 15.6 & 23.7 &75.0 & 23.3 & 35.6\\
Communication & 37.5 & 11.3 & 17.4 &43.8 & 13.2 & 20.3 &37.5 & 11.8 & 17.9 &47.1 & 15.7 & 23.5\\
\bottomrule
\textbf{Average} & \textbf{38.9} & \textbf{12.2} & \textbf{17.8} & \textbf{51.7} & \textbf{16.7} & \textbf{24.2} & \textbf{48.0} & \textbf{15.4} & \textbf{22.5} & \textbf{74.9} & \textbf{29.9} & \textbf{41.8}\\
\bottomrule

\end{tabular}
\caption{Model performance on \textsc{Guzman} dataset after all processing steps.}
\label{tab:GUZMAN_detail}
\end{table}


\subsection{\textsc{Shah-I} Dataset}

\begin{table}[H]
\centering
\small
\begin{tabular}{lcccccccccccc}
\toprule
\bf App & \multicolumn{3}{c}{\bf Exact Tokens} & \multicolumn{3}{c}{\bf Partial Tokens} & \multicolumn{3}{c}{\bf Exact Types} & \multicolumn{3}{c}{\bf Partial Types} \\
\bf Category & \bf Prec & \bf Rec & \bf F1 & \bf Prec & \bf Rec & \bf F1 & \bf Prec & \bf Rec & \bf F1 & \bf Prec & \bf Rec & \bf F1 \\
\midrule
\multicolumn{3}{l}{a) Pre-processing:} \\
\midrule
Game & 70.5 & 20.0 & 31.2 &80.8 & 22.9 & 35.7 &71.0 & 26.5 & 38.6 &88.2 & 36.1 & 51.3\\
Productivity & 61.0 & 37.1 & 46.1 &64.9 & 39.4 & 49.1 &55.1 & 21.3 & 30.7 &75.5 & 36.6 & 49.3\\
Travel & 69.4 & 42.7 & 52.9 &73.8 & 45.4 & 56.2 &54.1 & 21.7 & 31.0 &76.2 & 34.8 & 47.8\\
Photography & 85.5 & 68.1 & 75.8 &88.5 & 70.5 & 78.5 &66.7 & 28.1 & 39.5 &88.2 & 52.6 & 65.9\\
Social & 86.5 & 61.2 & 71.7 &88.1 & 62.3 & 73.0 &58.8 & 32.3 & 41.7 &72.2 & 41.9 & 53.1\\
Communication & 79.8 & 61.5 & 69.5 &89.4 & 68.9 & 77.8 &55.0 & 35.5 & 43.1 &80.0 & 51.6 & 62.7\\
\bottomrule
\textbf{Average} & \textbf{75.5} & \textbf{48.4} & \textbf{57.9} & \textbf{80.9} & \textbf{51.6} & \textbf{61.7} & \textbf{60.1} & \textbf{27.6} & \textbf{37.4} & \textbf{80.0} & \textbf{42.3} & \textbf{55.0}\\
\bottomrule

\midrule
\multicolumn{3}{l}{b) Simulation I:} \\
\midrule
Game & 68.9 & 21.8 & 33.2 &75.6 & 23.9 & 36.4 &69.2 & 22.8 & 34.3 &83.9 & 32.9 & 47.3\\
Productivity & 36.9 & 15.7 & 22.0 &42.3 & 18.0 & 25.3 &42.2 & 18.2 & 25.5 &62.8 & 30.7 & 41.3\\
Travel & 50.0 & 9.6 & 16.1 &61.5 & 11.9 & 19.9 &57.9 & 12.8 & 21.0 &77.3 & 19.8 & 31.5\\
Photography & 62.5 & 7.0 & 12.7 &62.5 & 7.0 & 12.7 &71.4 & 9.4 & 16.7 &85.7 & 22.6 & 35.8\\
Social & 38.5 & 10.8 & 16.8 &42.3 & 11.8 & 18.5 &62.5 & 15.9 & 25.3 &73.7 & 22.2 & 34.1\\
Communication & 0.0 & 0.0 & 0.0 &0.0 & 0.0 & 0.0 &0.0 & 0.0 & 0.0 &33.3 & 6.9 & 11.4\\
\bottomrule
\textbf{Average} & \textbf{42.8} & \textbf{10.8} & \textbf{16.8} & \textbf{47.4} & \textbf{12.1} & \textbf{18.8} & \textbf{50.5} & \textbf{13.2} & \textbf{20.5} & \textbf{69.5} & \textbf{22.5} & \textbf{33.6}\\
\bottomrule

\midrule
\multicolumn{3}{l}{c) Simulation II:} \\
\midrule
Game & 80.0 & 19.2 & 31.0 &86.7 & 20.8 & 33.5 &75.0 & 21.1 & 33.0 &88.0 & 31.0 & 45.8\\
Productivity & 46.5 & 15.4 & 23.2 &50.7 & 16.8 & 25.3 &48.1 & 14.9 & 22.8 &67.7 & 25.3 & 36.8\\
Travel & 57.9 & 10.2 & 17.3 &63.2 & 11.1 & 18.9 &57.1 & 9.5 & 16.3 &81.2 & 15.5 & 26.0\\
Photography & 71.4 & 8.2 & 14.7 &71.4 & 8.2 & 14.7 &71.4 & 10.4 & 18.2 &92.9 & 27.1 & 41.9\\
Social & 64.7 & 14.9 & 24.2 &76.5 & 17.6 & 28.6 &75.0 & 16.4 & 26.9 &92.9 & 23.6 & 37.7\\
Communication & 25.0 & 4.8 & 8.0 &25.0 & 4.8 & 8.0 &25.0 & 5.0 & 8.3 &50.0 & 10.0 & 16.7\\
\bottomrule
\textbf{Average} & \textbf{57.6} & \textbf{12.1} & \textbf{19.7} & \textbf{62.2} & \textbf{13.2} & \textbf{21.5} & \textbf{58.6} & \textbf{12.9} & \textbf{20.9} & \textbf{78.8} & \textbf{22.1} & \textbf{34.1}\\
\bottomrule

\midrule
\multicolumn{3}{l}{d) Simulation III-3:} \\
\midrule

Game & 83.3 & 16.8 & 28.0 &91.7 & 18.5 & 30.8 &78.6 & 16.4 & 27.2 &95.0 & 28.4 & 43.7\\
Productivity & 45.5 & 15.3 & 22.9 &49.1 & 16.6 & 24.8 &51.3 & 16.5 & 25.0 &71.1 & 26.4 & 38.6\\
Travel & 60.0 & 14.5 & 23.3 &65.0 & 15.7 & 25.2 &56.2 & 15.3 & 24.0 &78.9 & 25.4 & 38.5\\
Photography & 63.6 & 12.5 & 20.9 &63.6 & 12.5 & 20.9 &77.8 & 16.3 & 26.9 &100.0 & 34.9 & 51.7\\
Social & 45.5 & 15.9 & 23.5 &54.5 & 19.0 & 28.2 &64.3 & 20.5 & 31.0 &80.0 & 27.3 & 40.7\\
Communication & 33.3 & 5.9 & 10.0 &33.3 & 5.9 & 10.0 &33.3 & 6.2 & 10.5 &66.7 & 12.5 & 21.1\\
\bottomrule
\textbf{Average} & \textbf{55.2} & \textbf{13.5} & \textbf{21.4} & \textbf{59.5} & \textbf{14.7} & \textbf{23.3} & \textbf{60.2} & \textbf{15.2} & \textbf{24.1} & \textbf{82.0} & \textbf{25.8} & \textbf{39.0}\\
\bottomrule

\end{tabular}
\caption{Model performance on \textsc{Shah-I} dataset after all processing steps.}
\label{tab:SHAH-I_detail}
\end{table}


\subsection{\textsc{Shah-II} Dataset}

\begin{table}[H]
\centering
\small
\begin{tabular}{lcccccccccccc}
\toprule
\bf App & \multicolumn{3}{c}{\bf Exact Tokens} & \multicolumn{3}{c}{\bf Partial Tokens} & \multicolumn{3}{c}{\bf Exact Types} & \multicolumn{3}{c}{\bf Partial Types} \\
\bf Category & \bf Prec & \bf Rec & \bf F1 & \bf Prec & \bf Rec & \bf F1 & \bf Prec & \bf Rec & \bf F1 & \bf Prec & \bf Rec & \bf F1 \\
\midrule
\multicolumn{3}{l}{a) Pre-processing:} \\
\midrule
Game & 38.8 & 14.1 & 20.7 &46.9 & 17.0 & 25.0 &51.7 & 20.0 & 28.8 &64.5 & 26.7 & 37.7\\
Productivity & 31.9 & 11.4 & 16.7 &38.0 & 13.5 & 20.0 &43.0 & 14.4 & 21.6 &71.2 & 26.5 & 38.6\\
Travel & 40.9 & 10.9 & 17.2 &47.7 & 12.7 & 20.1 &46.7 & 12.7 & 20.0 &63.9 & 20.9 & 31.5\\
Photography & 33.3 & 9.0 & 14.1 &33.3 & 9.0 & 14.1 &50.0 & 7.5 & 13.1 &62.5 & 9.4 & 16.4\\
Social & 40.0 & 13.2 & 19.9 &42.2 & 14.0 & 21.0 &48.4 & 20.5 & 28.8 &59.4 & 26.0 & 36.2\\
Communication & 17.2 & 7.9 & 10.9 &20.7 & 9.5 & 13.0 &36.4 & 10.5 & 16.3 &63.6 & 18.4 & 28.6\\
\bottomrule
\textbf{Average} & \textbf{33.7} & \textbf{11.1} & \textbf{16.6} & \textbf{38.1} & \textbf{12.6} & \textbf{18.9} & \textbf{46.0} & \textbf{14.3} & \textbf{21.4} & \textbf{64.2} & \textbf{21.3} & \textbf{31.5}\\
\bottomrule
\midrule

\multicolumn{3}{l}{b) Simulation I:} \\
\midrule
Game & 48.7 & 19.6 & 27.9 &56.4 & 22.7 & 32.4 &50.0 & 15.3 & 23.4 &66.7 & 22.2 & 33.3\\
Productivity & 37.1 & 14.1 & 20.4 &42.1 & 15.9 & 23.1 &46.2 & 14.8 & 22.4 &69.4 & 23.4 & 35.0\\
Travel & 51.4 & 13.5 & 21.4 &60.0 & 15.8 & 25.0 &50.0 & 12.4 & 19.8 &65.5 & 18.1 & 28.4\\
Photography & 50.0 & 1.8 & 3.4 &50.0 & 1.8 & 3.4 &50.0 & 2.0 & 3.8 &100.0 & 4.0 & 7.7\\
Social & 44.4 & 19.2 & 26.8 &46.7 & 20.2 & 28.2 &46.9 & 21.1 & 29.1 &57.6 & 26.8 & 36.5\\
Communication & 37.5 & 7.5 & 12.5 &37.5 & 7.5 & 12.5 &50.0 & 8.6 & 14.6 &50.0 & 8.6 & 14.6\\
\bottomrule
\textbf{Average} & \textbf{44.9} & \textbf{12.6} & \textbf{18.7} & \textbf{48.8} & \textbf{14.0} & \textbf{20.8} & \textbf{48.9} & \textbf{12.4} & \textbf{18.8} & \textbf{68.2} & \textbf{17.2} & \textbf{25.9}\\
\bottomrule
\midrule
\multicolumn{3}{l}{c) Simulation II:} \\
\midrule
Game & 42.6 & 26.1 & 32.4 &48.1 & 29.5 & 36.6 &43.8 & 21.2 & 28.6 &58.1 & 27.3 & 37.1\\
Productivity & 44.0 & 14.8 & 22.2 &48.0 & 16.2 & 24.2 &49.3 & 13.7 & 21.5 &70.3 & 20.4 & 31.6\\
Travel & 46.9 & 13.2 & 20.5 &56.2 & 15.8 & 24.7 &44.4 & 12.1 & 19.0 &58.6 & 17.2 & 26.6\\
Photography & 33.3 & 2.4 & 4.4 &33.3 & 2.4 & 4.4 &33.3 & 2.6 & 4.8 &66.7 & 5.1 & 9.5\\
Social & 40.5 & 20.8 & 27.5 &43.2 & 22.2 & 29.4 &42.9 & 21.1 & 28.2 &57.1 & 28.1 & 37.6\\
Communication & 22.2 & 8.0 & 11.8 &22.2 & 8.0 & 11.8 &28.6 & 8.3 & 12.9 &28.6 & 8.3 & 12.9\\
\bottomrule
\textbf{Average} & \textbf{38.2} & \textbf{14.2} & \textbf{19.8} & \textbf{41.8} & \textbf{15.7} & \textbf{21.9} & \textbf{40.4} & \textbf{13.2} & \textbf{19.2} & \textbf{56.6} & \textbf{17.7} & \textbf{25.9}\\
\bottomrule
\midrule

\multicolumn{3}{l}{d) Simulation III-3:} \\
\midrule
Game & 45.2 & 22.6 & 30.2 &54.8 & 27.4 & 36.5 &44.0 & 17.7 & 25.3 &70.8 & 27.4 & 39.5\\
Productivity & 50.0 & 16.1 & 24.4 &54.4 & 17.5 & 26.5 &56.8 & 14.8 & 23.5 &70.8 & 20.1 & 31.3\\
Travel & 61.9 & 16.0 & 25.5 &71.4 & 18.5 & 29.4 &60.0 & 13.6 & 22.2 &76.5 & 19.7 & 31.3\\
Photography & 66.7 & 5.9 & 10.8 &66.7 & 5.9 & 10.8 &66.7 & 6.5 & 11.8 &66.7 & 6.5 & 11.8\\
Social & 52.6 & 17.5 & 26.3 &57.9 & 19.3 & 28.9 &69.2 & 21.4 & 32.7 &73.3 & 26.2 & 38.6\\
Communication & 25.0 & 5.3 & 8.7 &25.0 & 5.3 & 8.7 &50.0 & 11.1 & 18.2 &50.0 & 11.1 & 18.2\\
\bottomrule
\textbf{Average} & \textbf{50.2} & \textbf{13.9} & \textbf{21.0} & \textbf{55.0} & \textbf{15.6} & \textbf{23.5} & \textbf{57.8} & \textbf{14.2} & \textbf{22.3} & \textbf{68.0} & \textbf{18.5} & \textbf{28.4}\\
\bottomrule
\midrule
\end{tabular}
\caption{Model performance on \textsc{Shah-II} dataset after all processing steps.}
\label{tab:SHAH-II_detail}
\end{table}


\subsection{\textsc{Sanger} Dataset}

\begin{table}[H]
\centering
\small
\begin{tabular}{lcccccccccccc}
\toprule
\bf App & \multicolumn{3}{c}{\bf Exact Tokens} & \multicolumn{3}{c}{\bf Partial Tokens} & \multicolumn{3}{c}{\bf Exact Types} & \multicolumn{3}{c}{\bf Partial Types} \\
\bf Category & \bf Prec & \bf Rec & \bf F1 & \bf Prec & \bf Rec & \bf F1 & \bf Prec & \bf Rec & \bf F1 & \bf Prec & \bf Rec & \bf F1 \\
\midrule
\multicolumn{3}{l}{a) Pre-processing:} \\
\midrule
Weather Apps & 72.9 & 62.1 & 67.1 &80.6 & 68.6 & 74.1 &68.3 & 51.2 & 58.5 &78.8 & 61.9 & 69.3\\
Sport News & 69.2 & 50.3 & 58.2 &76.7 & 55.8 & 64.6 &56.8 & 27.8 & 37.3 &73.5 & 40.0 & 51.8\\
Social Networks & 71.8 & 56.0 & 62.9 &77.6 & 60.6 & 68.0 &61.9 & 47.3 & 53.6 &72.7 & 58.2 & 64.6\\
Office Tools & 66.7 & 45.2 & 53.9 &73.0 & 49.5 & 59.0 &61.3 & 36.8 & 46.0 &75.3 & 46.4 & 57.4\\
News Apps & 63.7 & 53.5 & 58.2 &69.5 & 58.4 & 63.5 &57.5 & 44.6 & 50.3 &65.9 & 53.6 & 59.1\\
Navigation Apps & 63.9 & 57.2 & 60.4 &69.0 & 61.8 & 65.2 &54.1 & 45.5 & 49.5 &63.6 & 55.4 & 59.3\\
Music Players & 69.6 & 41.8 & 52.3 &72.0 & 43.3 & 54.1 &61.4 & 33.3 & 43.2 &71.6 & 45.7 & 55.8\\
Instant Messengers & 68.9 & 49.0 & 57.3 &81.1 & 57.7 & 67.5 &64.4 & 32.2 & 43.0 &81.5 & 48.9 & 61.1\\
Games & 73.8 & 31.4 & 44.0 &76.9 & 32.7 & 45.9 &68.8 & 30.1 & 41.9 &76.9 & 41.1 & 53.6\\
Fitness Tracker & 81.1 & 52.9 & 64.1 &86.1 & 56.1 & 68.0 &74.0 & 34.9 & 47.4 &81.8 & 42.5 & 55.9\\
Alarm Clocks & 72.1 & 49.5 & 58.7 &78.3 & 53.7 & 63.7 &70.4 & 43.7 & 53.9 &80.4 & 51.7 & 62.9\\
\bottomrule
\textbf{Average} & \textbf{70.3} & \textbf{49.9} & \textbf{57.9} & \textbf{76.4} & \textbf{54.4} & \textbf{63.1} & \textbf{63.5} & \textbf{38.9} & \textbf{47.7} & \textbf{74.7} & \textbf{49.6} & \textbf{59.2}\\
\bottomrule
\midrule

\multicolumn{3}{l}{b) Simulation I:} \\
\midrule
Weather Apps & 54.1 & 38.5 & 44.9 &63.5 & 45.2 & 52.8 &61.5 & 41.0 & 49.2 &72.2 & 50.0 & 59.1\\
Sport News & 56.8 & 34.7 & 43.1 &66.2 & 40.5 & 50.3 &46.8 & 26.2 & 33.6 &62.0 & 36.9 & 46.3\\
Social Networks & 57.1 & 43.8 & 49.6 &64.3 & 49.3 & 55.8 &48.6 & 32.7 & 39.1 &61.1 & 42.3 & 50.0\\
Office Tools & 51.3 & 25.3 & 33.9 &57.9 & 28.6 & 38.3 &54.1 & 27.7 & 36.7 &68.8 & 37.0 & 48.1\\
News Apps & 55.0 & 32.0 & 40.4 &60.0 & 34.9 & 44.1 &57.4 & 36.4 & 44.6 &66.2 & 45.8 & 54.1\\
Navigation Apps & 57.4 & 40.3 & 47.4 &62.8 & 44.0 & 51.8 &53.7 & 38.3 & 44.7 &65.3 & 50.0 & 56.6\\
Music Players & 60.3 & 29.7 & 39.8 &64.7 & 31.9 & 42.7 &57.4 & 27.6 & 37.2 &64.6 & 31.6 & 42.5\\
Instant Messengers & 66.7 & 27.7 & 39.2 &78.6 & 32.7 & 46.2 &60.0 & 19.5 & 29.4 &81.5 & 28.6 & 42.3\\
Games & 69.7 & 23.7 & 35.4 &69.7 & 23.7 & 35.4 &86.4 & 27.1 & 41.3 &88.9 & 34.3 & 49.5\\
Fitness Tracker & 55.9 & 28.9 & 38.2 &64.4 & 33.3 & 43.9 &68.4 & 39.4 & 50.0 &78.7 & 48.5 & 60.0\\
Alarm Clocks & 63.5 & 35.7 & 45.7 &66.7 & 37.5 & 48.0 &75.0 & 37.5 & 50.0 &81.0 & 42.5 & 55.7\\
\bottomrule
\textbf{Average} & \textbf{58.9} & \textbf{32.8} & \textbf{41.6} & \textbf{65.3} & \textbf{36.5} & \textbf{46.3} & \textbf{60.8} & \textbf{32.1} & \textbf{41.4} & \textbf{71.8} & \textbf{40.7} & \textbf{51.3}\\
\bottomrule
\midrule

\multicolumn{3}{l}{c) Simulation II:} \\
\midrule
Weather Apps & 51.4 & 40.9 & 45.5 &60.8 & 48.4 & 53.9 &60.8 & 44.9 & 51.7 &68.5 & 53.6 & 60.2\\
Sport News & 60.8 & 34.1 & 43.7 &66.7 & 37.4 & 47.9 &55.2 & 24.2 & 33.7 &66.7 & 33.3 & 44.4\\
Social Networks & 63.9 & 46.9 & 54.1 &66.7 & 49.0 & 56.5 &64.0 & 43.2 & 51.6 &72.0 & 48.6 & 58.1\\
Office Tools & 50.0 & 26.2 & 34.3 &57.4 & 30.0 & 39.4 &54.5 & 28.6 & 37.5 &70.2 & 38.1 & 49.4\\
News Apps & 50.9 & 36.4 & 42.5 &56.5 & 40.4 & 47.1 &48.1 & 38.1 & 42.5 &57.8 & 49.5 & 53.3\\
Navigation Apps & 48.1 & 33.6 & 39.6 &56.8 & 39.7 & 46.7 &46.9 & 35.7 & 40.5 &59.4 & 45.2 & 51.4\\
Music Players & 50.0 & 23.6 & 32.1 &55.8 & 26.4 & 35.8 &48.8 & 25.3 & 33.3 &59.1 & 31.3 & 40.9\\
Instant Messengers & 46.9 & 23.1 & 30.9 &56.2 & 27.7 & 37.1 &39.1 & 17.0 & 23.7 &60.0 & 28.3 & 38.5\\
Games & 48.6 & 25.4 & 33.3 &51.4 & 26.9 & 35.3 &64.0 & 32.7 & 43.2 &74.1 & 40.8 & 52.6\\
Fitness Tracker & 52.4 & 23.4 & 32.4 &59.5 & 26.6 & 36.8 &63.4 & 32.5 & 43.0 &72.7 & 40.0 & 51.6\\
Alarm Clocks & 55.3 & 33.3 & 41.6 &61.7 & 37.2 & 46.4 &58.1 & 30.0 & 39.6 &71.4 & 41.7 & 52.6\\
\bottomrule
\textbf{Average} & \textbf{52.6} & \textbf{31.5} & \textbf{39.1} & \textbf{59.0} & \textbf{35.4} & \textbf{43.9} & \textbf{54.8} & \textbf{32.0} & \textbf{40.0} & \textbf{66.5} & \textbf{40.9} & \textbf{50.3}\\
\bottomrule

\midrule

\multicolumn{3}{l}{d) Simulation III-3:} \\
\midrule
Weather Apps & 52.8 & 42.2 & 46.9 &61.1 & 48.9 & 54.3 &62.5 & 45.5 & 52.6 &68.6 & 53.0 & 59.8\\
Sport News & 64.6 & 34.8 & 45.3 &70.8 & 38.2 & 49.6 &61.5 & 25.0 & 35.6 &72.4 & 32.8 & 45.2\\
Social Networks & 64.5 & 42.6 & 51.3 &67.7 & 44.7 & 53.8 &66.7 & 40.0 & 50.0 &76.2 & 45.7 & 57.1\\
Office Tools & 46.3 & 24.6 & 32.1 &56.7 & 30.2 & 39.4 &51.8 & 28.7 & 36.9 &70.2 & 39.6 & 50.6\\
News Apps & 54.6 & 36.1 & 43.4 &57.7 & 38.1 & 45.9 &53.0 & 37.6 & 44.0 &59.7 & 46.2 & 52.1\\
Navigation Apps & 49.3 & 28.9 & 36.5 &58.2 & 34.2 & 43.1 &46.9 & 28.0 & 35.1 &63.5 & 40.2 & 49.3\\
Music Players & 55.8 & 22.0 & 31.6 &58.1 & 22.9 & 32.9 &55.9 & 23.2 & 32.8 &63.2 & 29.3 & 40.0\\
Instant Messengers & 53.1 & 27.4 & 36.2 &59.4 & 30.6 & 40.4 &47.8 & 22.0 & 30.1 &62.5 & 30.0 & 40.5\\
Games & 45.5 & 22.4 & 30.0 &48.5 & 23.9 & 32.0 &60.0 & 30.6 & 40.5 &70.4 & 38.8 & 50.0\\
Fitness Tracker & 58.1 & 20.2 & 30.0 &61.3 & 21.3 & 31.7 &67.7 & 28.0 & 39.6 &76.5 & 34.7 & 47.7\\
Alarm Clocks & 54.5 & 33.3 & 41.4 &61.4 & 37.5 & 46.6 &55.6 & 27.8 & 37.0 &70.0 & 38.9 & 50.0\\
\bottomrule
\textbf{Average} & \textbf{54.5} & \textbf{30.4} & \textbf{38.6} & \textbf{60.1} & \textbf{33.7} & \textbf{42.7} & \textbf{57.2} & \textbf{30.6} & \textbf{39.5} & \textbf{68.5} & \textbf{39.0} & \textbf{49.3}\\
\bottomrule

\end{tabular}
\caption{Model performance on \textsc{Sanger} dataset after all processing steps.}
\label{tab:SANGER_detail}

\end{table}

%% file: appendix_B.tex
\section{Detailed results of the training procedure \textsc{AppCat}}

\begin{table}[H]
\centering
\small
\begin{tabular}{lcccccccccccc}
\toprule
\bf App & \multicolumn{3}{c}{\bf Exact Tokens} & \multicolumn{3}{c}{\bf Partial Tokens} & \multicolumn{3}{c}{\bf Exact Types} & \multicolumn{3}{c}{\bf Partial Types} \\
\bf Category & \bf Prec & \bf Rec & \bf F1 & \bf Prec & \bf Rec & \bf F1 & \bf Prec & \bf Rec & \bf F1 & \bf Prec & \bf Rec & \bf F1 \\
\midrule
\multicolumn{3}{l}{a) \textsc{GUZMAN}:} \\
\midrule
Game & 56.4 & 28.2 & 37.6 &69.0 & 34.6 & 46.1 &69.1 & 31.2 & 43.0 &84.1 & 43.0 & 56.9\\
Productivity & 40.1 & 16.7 & 23.6 &53.4 & 21.7 & 30.9 &47.7 & 19.5 & 27.6 &73.4 & 32.9 & 45.4\\
Travel & 44.3 & 18.7 & 26.3 &53.0 & 22.4 & 31.5 &57.4 & 22.3 & 32.1 &80.5 & 31.8 & 45.6\\
Photography & 31.8 & 10.9 & 16.2 &45.1 & 14.5 & 21.9 &39.5 & 12.3 & 18.8 &71.7 & 20.6 & 32.0\\
Social & 44.5 & 10.0 & 16.4 &49.3 & 11.7 & 18.9 &50.0 & 12.0 & 19.4 &55.8 & 14.6 & 23.2\\
Communication & 0.0 & 0.0 & 0.0 &5.0 & 1.7 & 2.5 &0.0 & 0.0 & 0.0 &5.0 & 1.7 & 2.5\\
\bottomrule
\textbf{Average} & \textbf{36.2} & \textbf{14.1} & \textbf{20.0} & \textbf{45.8} & \textbf{17.8} & \textbf{25.3} & \textbf{43.9} & \textbf{16.2} & \textbf{23.5} & \textbf{61.8} & \textbf{24.1} & \textbf{34.3}\\
\bottomrule

\midrule
\multicolumn{3}{l}{b) \textsc{shah-I}:} \\
\midrule
Game & 77.2 & 29.2 & 42.3 &83.7 & 31.8 & 46.1 &80.8 & 32.3 & 46.2 &89.2 & 43.7 & 58.7\\
Productivity & 71.2 & 20.5 & 31.8 &76.8 & 22.2 & 34.5 &76.2 & 20.3 & 32.0 &86.9 & 27.7 & 42.0\\
Travel & 54.2 & 11.7 & 19.3 &54.2 & 11.7 & 19.3 &60.0 & 15.0 & 24.0 &61.7 & 18.5 & 28.5\\
Photography & 47.5 & 15.6 & 23.5 &57.5 & 20.6 & 30.4 &57.5 & 18.0 & 27.4 &69.2 & 26.3 & 38.1\\
Social & 10.0 & 1.1 & 2.0 &10.0 & 1.1 & 2.0 &10.0 & 1.1 & 2.0 &10.0 & 1.1 & 2.0\\
Communication & 0.0 & 0.0 & 0.0 &0.0 & 0.0 & 0.0 &0.0 & 0.0 & 0.0 &0.0 & 0.0 & 0.0\\
\bottomrule
\textbf{Average} & \textbf{43.4} & \textbf{13.0} & \textbf{19.8} & \textbf{47.0} & \textbf{14.6} & \textbf{22.0} & \textbf{47.4} & \textbf{14.4} & \textbf{21.9} & \textbf{52.8} & \textbf{19.6} & \textbf{28.2}\\
\bottomrule

\midrule
\multicolumn{3}{l}{c) \textsc{shah-II}:} \\
\midrule
Game & 43.8 & 17.7 & 25.2 &48.6 & 22.1 & 30.4 &67.0 & 26.0 & 37.5 &77.5 & 29.6 & 42.9\\
Productivity & 37.1 & 10.5 & 16.3 &42.5 & 12.1 & 18.8 &45.7 & 13.0 & 20.2 &56.6 & 17.4 & 26.6\\
Travel & 28.3 & 6.0 & 9.9 &38.3 & 7.1 & 12.0 &28.3 & 6.4 & 10.5 &38.3 & 7.5 & 12.6\\
Photography & 0.0 & 0.0 & 0.0 &0.0 & 0.0 & 0.0 &0.0 & 0.0 & 0.0 &0.0 & 0.0 & 0.0\\
Social & 0.0 & 0.0 & 0.0 &0.0 & 0.0 & 0.0 &5.0 & 1.1 & 1.8 &5.0 & 1.1 & 1.8\\
Communication & 0.0 & 0.0 & 0.0 &0.0 & 0.0 & 0.0 &0.0 & 0.0 & 0.0 &0.0 & 0.0 & 0.0\\
\bottomrule
\textbf{Average} & \textbf{18.2} & \textbf{5.7} & \textbf{8.6} & \textbf{21.6} & \textbf{6.9} & \textbf{10.2} & \textbf{24.3} & \textbf{7.8} & \textbf{11.7} & \textbf{29.6} & \textbf{9.3} & \textbf{14.0}\\
\bottomrule
\midrule
\multicolumn{3}{l}{d) \textsc{sanger}:} \\
\midrule
Weather Apps & 72.2 & 36.9 & 48.8 &76.7 & 40.7 & 53.1 &72.2 & 38.7 & 50.4 &78.1 & 44.0 & 56.3\\
Sport News & 54.4 & 22.4 & 31.7 &60.8 & 25.2 & 35.6 &58.8 & 22.0 & 32.1 &65.5 & 25.0 & 36.2\\
Social Networks & 92.5 & 37.0 & 52.8 &92.5 & 37.0 & 52.8 &98.0 & 40.2 & 57.0 &98.0 & 40.2 & 57.0\\
Office Tools & 37.1 & 14.3 & 20.7 &45.9 & 17.1 & 24.9 &40.0 & 16.8 & 23.6 &53.6 & 23.3 & 32.5\\
News Apps & 57.4 & 18.5 & 28.0 &61.9 & 20.5 & 30.7 &64.2 & 21.4 & 32.2 &68.1 & 23.6 & 35.0\\
Navigation Apps & 59.3 & 30.2 & 40.1 &65.3 & 33.5 & 44.3 &61.4 & 31.6 & 41.7 &67.3 & 35.2 & 46.2\\
Music Players & 46.3 & 14.9 & 22.6 &53.3 & 16.5 & 25.2 &50.2 & 15.9 & 24.2 &55.7 & 17.5 & 26.6\\
Instant Messengers & 31.7 & 8.3 & 13.1 &51.7 & 11.9 & 19.4 &41.7 & 12.6 & 19.3 &51.7 & 16.3 & 24.7\\
Games & 65.0 & 14.1 & 23.1 &70.0 & 16.1 & 26.1 &65.0 & 14.8 & 24.1 &70.0 & 16.8 & 27.1\\
Fitness Tracker & 31.2 & 12.6 & 17.9 &46.2 & 18.6 & 26.5 &30.4 & 12.1 & 17.3 &45.4 & 18.3 & 26.1\\
Alarm Clocks & 72.7 & 22.2 & 34.0 &72.7 & 22.2 & 34.0 &71.7 & 21.5 & 33.0 &74.2 & 25.7 & 38.2\\
\bottomrule
\textbf{Average} & \textbf{56.3} & \textbf{21.0} & \textbf{30.3} & \textbf{63.4} & \textbf{23.6} & \textbf{33.9} & \textbf{59.4} & \textbf{22.5} & \textbf{32.3} & \textbf{66.1} & \textbf{26.0} & \textbf{36.9}\\
\bottomrule

\end{tabular}
\caption{Model performance on datasets for the training procedure \textsc{AppCat}.}
\label{tab:AppCat}
\end{table}

%% file: appendix_C.tex
\section{Detailed results of the training procedure \textsc{CCV-Ext}}

\begin{table}[H]
\centering
\small
\begin{tabular}{lcccccccccccc}
\toprule
\bf App & \multicolumn{3}{c}{\bf Exact Tokens} & \multicolumn{3}{c}{\bf Partial Tokens} & \multicolumn{3}{c}{\bf Exact Types} & \multicolumn{3}{c}{\bf Partial Types} \\
\bf Category & \bf Prec & \bf Rec & \bf F1 & \bf Prec & \bf Rec & \bf F1 & \bf Prec & \bf Rec & \bf F1 & \bf Prec & \bf Rec & \bf F1 \\
\midrule
\multicolumn{3}{l}{a) \textsc{GUZMAN}:} \\
\midrule
Game & 27.5 & 16.2 & 20.4 &35.6 & 20.9 & 26.4 &42.9 & 25.3 & 31.8 &71.0 & 57.1 & 63.3\\
Productivity & 20.3 & 28.9 & 23.8 &27.8 & 39.6 & 32.7 &31.1 & 35.0 & 32.9 &56.5 & 64.3 & 60.2\\
Travel & 21.2 & 11.5 & 14.9 &30.8 & 16.7 & 21.6 &31.7 & 16.8 & 22.0 &60.0 & 39.5 & 47.6\\
Photography & 29.2 & 20.0 & 23.8 &44.6 & 30.5 & 36.2 &29.6 & 23.9 & 26.4 &53.2 & 37.3 & 43.9\\
Social & 27.8 & 18.5 & 22.2 &38.0 & 25.2 & 30.3 &29.5 & 20.0 & 23.8 &49.2 & 35.6 & 41.3\\
Communication & 18.2 & 11.1 & 13.8 &30.3 & 18.5 & 23.0 &21.9 & 13.5 & 16.7 &42.4 & 26.9 & 32.9\\
\bottomrule
\textbf{Average} & \textbf{24.0} & \textbf{17.7} & \textbf{19.8} & \textbf{34.5} & \textbf{25.2} & \textbf{28.4} & \textbf{31.1} & \textbf{22.4} & \textbf{25.6} & \textbf{55.4} & \textbf{43.4} & \textbf{48.2}\\
\bottomrule
\midrule
\multicolumn{3}{l}{b) \textsc{shah-I}:} \\
\midrule
Game & 24.8 & 22.7 & 23.7 &26.6 & 24.4 & 25.4 &31.5 & 34.3 & 32.9 &50.6 & 59.7 & 54.8\\
Productivity & 15.0 & 33.3 & 20.7 &16.7 & 37.1 & 23.0 &20.1 & 39.8 & 26.7 &34.5 & 64.4 & 45.0\\
Travel & 14.3 & 19.3 & 16.4 &16.1 & 21.7 & 18.5 &19.4 & 22.0 & 20.6 &36.4 & 40.7 & 38.4\\
Photography & 26.4 & 25.0 & 25.7 &34.0 & 32.1 & 33.0 &30.4 & 32.6 & 31.5 &48.0 & 55.8 & 51.6\\
Social & 19.7 & 19.0 & 19.4 &26.2 & 25.4 & 25.8 &33.3 & 36.4 & 34.8 &48.0 & 54.5 & 51.1\\
Communication & 7.4 & 11.8 & 9.1 &11.1 & 17.6 & 13.6 &12.5 & 18.8 & 15.0 &26.1 & 37.5 & 30.8\\
\bottomrule
\textbf{Average} & \textbf{17.9} & \textbf{21.8} & \textbf{19.2} & \textbf{21.8} & \textbf{26.4} & \textbf{23.2} & \textbf{24.5} & \textbf{30.7} & \textbf{26.9} & \textbf{40.6} & \textbf{52.1} & \textbf{45.3}\\
\bottomrule
\midrule
\multicolumn{3}{l}{c) \textsc{shah-II}:} \\
\midrule
Game & 18.4 & 21.7 & 19.9 &21.4 & 25.3 & 23.2 &26.6 & 27.9 & 27.2 &46.4 & 52.5 & 49.2\\
Productivity & 19.8 & 29.3 & 23.6 &23.7 & 35.1 & 28.3 &29.0 & 36.5 & 32.4 &45.3 & 55.1 & 49.7\\
Travel & 17.6 & 26.2 & 21.1 &21.0 & 31.2 & 25.1 &23.0 & 26.2 & 24.5 &40.3 & 44.6 & 42.3\\
Photography & 15.2 & 20.6 & 17.5 &21.7 & 29.4 & 25.0 &19.0 & 25.8 & 21.9 &34.1 & 48.4 & 40.0\\
Social & 21.2 & 19.3 & 20.2 &25.0 & 22.8 & 23.9 &22.2 & 19.0 & 20.5 &31.6 & 28.6 & 30.0\\
Communication & 8.3 & 11.1 & 9.5 &8.3 & 11.1 & 9.5 &9.5 & 11.8 & 10.5 &10.0 & 11.8 & 10.8\\
\bottomrule
\textbf{Average} & \textbf{16.8} & \textbf{21.4} & \textbf{18.6} & \textbf{20.2} & \textbf{25.8} & \textbf{22.5} & \textbf{21.6} & \textbf{24.5} & \textbf{22.8} & \textbf{34.6} & \textbf{40.2} & \textbf{37.0}\\
\bottomrule

\end{tabular}
\caption{Model performance on datasets for the training procedure \textsc{CCV-Ext}.}
\label{tab:CCV-Ext}
\end{table}